\documentclass[oneside,11pt]{article}

\usepackage[T1]{fontenc} 
\usepackage[english]{babel}    
\usepackage{enumerate} 
\usepackage{graphics,latexsym,amsfonts}      
\usepackage{graphicx}   
\usepackage{amssymb,amsthm,hyperref,mathtools}   
\usepackage{mathrsfs} 
\usepackage{float}
\usepackage[dvipsnames]{xcolor} 
\usepackage{picture}
\usepackage{ulem}    
\hypersetup{
    colorlinks, 
    linkcolor={red!50!black},  
    citecolor={blue!50!black},  
    urlcolor={blue!80!black}  
}
\usepackage[longnamesfirst]{natbib}   
\usepackage{tabularx}

\def\reference#1{\href{#1}{Cliquer ici pour voir une r\'ef\'erence.}} 
\usepackage{pstricks} 
\usepackage{hyperref}

\usepackage{caption}
\usepackage{sectsty} 

\usepackage{bigfoot} % to allow verbatim in footnote
\usepackage[numbered,framed]{matlab-prettifier}%matlab
\usepackage{filecontents}%matlab

\lstMakeShortInline"

\usepackage{setspace}
\usepackage{caption}
\sectionfont{\normalfont\scshape\centering}
\subsectionfont{\Large\normalfont\centering}
\providecommand{\U}[1]{\protect \rule{.1in}{.1in}}

\usepackage[top=1.1in,bottom=1.1in,left=1.25in]{geometry}
\def\reference#1{\href{#1}{Click to see a reference}} 
\def\u{\underline}
\def\r{\rightarrowtail}
\def\B{\,\mathbf{B}\,}

\def\vs{\vspace{-0,1cm}}

\def\L{\mathscr{L}}
\def\X{\mathscr{X}}
\def\es{\varnothing}
\makeatletter

\newtheoremstyle{mytheoremstyle} % name
    {\topsep}                    % Space above
    {\topsep}                    % Space below
    {\itshape}                   % Body font
    {}                           % Indent amount
    {\scshape}                   % Theorem head font
    {.}                          % Punctuation after theorem head
    {.5em}                       % Space after theorem head
    {}  % Theorem head spec (can be left empty, meaning 'normal')

\theoremstyle{mytheoremstyle}
\newtheorem{theorem}{Theorem} % reset theorem numbering for each chapter
\newtheorem*{theorem*}{Theorem}

\newtheorem{lemma}{Lemma}
 
\newtheorem*{corollary*}{Corollary} 
\newtheoremstyle{mydefinitionstyle} % name
    {\topsep}                    % Space above
    {\topsep}                    % Space below
    {}                   % Body font
    {}                           % Indent amount
    {\scshape}                   % Theorem head font
    {.}                          % Punctuation after theorem head
    {.5em}                       % Space after theorem head
    {}  % Theorem head spec (can be left empty, meaning "normal"ï)
\theoremstyle{mydefinitionstyle}
\newtheorem{definition}{Definition} % definition numbers are dependent on theorem numbers
\newtheorem*{question*}{Question}

\makeatletter

\title{On the number of non-isomorphic choices on four elements\thanks{The authors thank Davide Carpentiere and Marco Mariotti for several comments. Alfio Giarlotta acknowledges the support of ``Ministero dell'Istruzione, dell'Universit\`a e della Ricerca (MIUR) -- PRIN 2017'', project \textit{Multiple Criteria Decision Analysis and Multiple Criteria Decision Theory}, grant 2017CY2NCA.}}

\author{\textsc{Alfio Giarlotta}\thanks{Department of Economics and Business, University of Catania, Catania, Italy. alfio.giarlotta@unict.it}, 
\textsc{Angelo Petralia}\thanks{University of Turin \& Collegio Carlo Alberto, Turin, Italy. angelo.petralia@carloalberto.org (corresponding author)},  
\textsc{Stephen Watson}\thanks{Department\:of\:Mathematics\:and\:Statistics,\:York\:University,\:Toronto,\:Canada.\;watson@mathstat.yorku.ca}} 

\usepackage{fancyhdr} 
\date{}

\begin{document} 
	\maketitle

\vs\vs
\begin{abstract}
\noindent
We use a combinatorial approach to compute the number of non-isomorphic choices on four elements that can be explained by models of bounded rationality.
%\begin{itemize}
%	\item This computation allows the application of an algorithm to estimate the fraction of choices justifiable by these models.
%	\item Our method can be extended to evaluate other (existing or future) models of bounded rationality.
%\end{itemize}
%To that end, we apply a uniform combinatorial method.
%This computation allows the application of an algorithm to estimate the fraction of choices justifiable by these models. 
%Our approach can be extended to evaluate other (existing or future) models of bounded rationality.\smallskip
\smallskip



\end{abstract}

%%%%%%%%%%%%%%%%%%%%%%%%%%%%%%%%%
%%%%%%%%%%%%%%%%%%%%%%%%%%%%%%%%%
\section{\textbf{Specification table}}
\vs\vs\vs
\begin{table}[H]
\begin{center}
\footnotesize 
\begin{tabular}{|l|l|}
\hline
Subject area & Economics\\
More specific subject area & Microeconomic Theory, Behavioral Economics\\
Method name & Computing the number of non-isomorphic choices on four items\\
Original method and references & \cite{GiarlottaPetraliaWatson2021}\\
Resource availability & \href{https://it.mathworks.com/products/matlab.html}{https://it.mathworks.com/products/matlab.html}\\
\hline
\end{tabular}
\end{center}
\end{table}

%%%%%%%%%%%%%%%%%%%%%%%%%%%%%%%%%
%%%%%%%%%%%%%%%%%%%%%%%%%%%%%%%%%

\section{\textbf{Motivation}}

The notion of \textsl{rationalizability} pioneered by \cite{Samuelson1938} identifies a narrow kind of rational choice behavior.  
Starting from the seminal work of~\cite{Simon1955}, rationalizability has been weakened by the notion of \textsl{bounded rationality}, which allows to explain a larger fraction of choices by more flexible paradigms.  
In view of applications, it is interesting to compare existing bounded rationality models according to the fraction of choices justifiable by them.  
To that end, in this note we give a detailed proof of a related result, namely Lemma~8 in \cite{GiarlottaPetraliaWatson2021}. 

Specifically, we determine --up to relabelings of alternatives (i.e., \textsl{up to isomorphisms})-- the number of choices on four elements that can be  explained by several well-known models of bounded rationality. 
This is the key numerical input for an algorithm, which establishes an upper bound to the fraction of choices that are boundedly rationalizable by any of these models. 
The relevance of our computations lies in the fact that the combinatorial approach developed by \citet{GiarlottaPetraliaWatson2021} applies to any --existing or future-- model of bounded rationality, as long as one can determine the number of non-isomorphic choices on four items explained by it.
%Therefore, the possible relevance of the result proved in this note becomes apparent.  

%%%%%%%%%%%%%%%%%%%%%%%%%%%%%%%%%
%%%%%%%%%%%%%%%%%%%%%%%%%%%%%%%%%

\section{\textbf{Method background}} \label{SECT:background}

Let $X$ be a nonempty finite set of options, called the \textsl{ground set}.
Any nonempty set $A \subseteq X$ is a \textsl{menu}, and $\X = 2^X \setminus \{\es\}$ is the family of all menus.
Elements of menus are also called \textsl{items}. 
A \textsl{choice function} (for short, a \textsl{choice}) on $X$ is a map $c \colon \X\rightarrow X$ such that $c(A)\in A$ for any $A\in\X$. 
%
%\begin{definition} \label{DEF:isomorphic_choices}
%	Two choices $c \colon \X \to X$ and $c' \colon \X' \to X'$ are \textsl{isomorphic} if there exists a bijection $\sigma \colon X \to X'$ such that $\sigma(c(A)) = c'(\sigma(A))$ for any $A\in\X$. 
%	A \textsl{property} $\mathscr{P}$ \textsl{of choices} is a nonempty proper subset of choices that is closed under isomorphism.	
%\end{definition}
%
The properties of choices that we discuss in this note are listed below, along with some additional models of bounded rationality that are equivalent to them.\footnote{{ }Models are listed in the same order as in the main result of this paper, namely Theorem~\ref{THM:estimates}.}\vs

\begin{description}
	\item[$\bullet$ \textsf{Status quo bias (SQB)} \citep{ApesteguiaBallester2013}:] By definition, $c$ is SQB iff it is either \textsl{extreme status quo bias} (ESQB) or \textsl{weak status quo bias} (WSQB).\vs\vs
	\begin{description}
		\item[\textsf{ESQB}:] There exists a triple $(\rhd,z,Q)$, where $\rhd$ is a linear order on $X$, $ z$ is a selected item of $X$, and $Q \subseteq \{x \in X : x \rhd z\}$, such that for any $S \in \X$,\vs\vs
		\begin{itemize}
			\item[(1)] if $z \notin S$, then $c(S)=\max(S,\rhd)\,$,\vs 
			\item[(2)] if $z \in S$ and $Q \cap S = \es$, then $c(S)=z\,$, and\vs  
			\item[(3)] if $z \in S$ and $Q \cap S \neq \es$, then $c(S) = \max(Q \cap S,\rhd)\,$.\vs
		\end{itemize}
		\item[\textsf{WSQB}:] There exists a triple $(\rhd,z,Q)$, where $\rhd$ is a linear order on $X$, $z$ is a selected item of $X$, and $ Q \subseteq \{x \in X : x \rhd z\}$, such that for any $S \in \X$,\vs\vs 
		\begin{itemize}
			\item[(1)] if $z \notin S$, then $c(S)=\max(S,\rhd)\,$,\vs 
			\item[(2)] if $z \in S$ and $Q \cap S = \es$, then $c(S)=z\,$, and\vs  
			\item[(3$'$)] if $z \in S$ and $Q \cap S \neq \es$, then $c(S) = \max(S\setminus \{z\},\rhd)\,$.
		\end{itemize}\vs
	\end{description} 
	
	\item[$\bullet$ \textsf{List rational (LR)} \citep{Yildiz2016}:] By definition, $c$ is LR iff there is a linear order $\rhd$ on $X$ (a \textsl{list}) such that for any $A\in\X$ of size at least two, the equality $c(A)=c(\{c(A\setminus{x}),x\})$ holds, where $x=\min(A,\rhd)$.%\footnote{{ }$\min(A,\rhd)$ is the unique item $x$ such that =\{x \in X : x \,\rhd y\, \text{ for no } y \in A\}$ is the set of \textsl{minimal elements} of $A$.} 
	%the relation $\textbf{f}_c$ of \textsl{revealed-to-follow}, defined by $x\,\textbf{f}_c\,y$ if for some $A\in\X$, either (i) $x=c(A\cup y)$ and $[y=c(\{x,y\})$ or $x\neq c(A)]$, or (ii) $x\neq c(A\cup y)$ and $[x=c(\{x,y\})$ or $x=c(A)]$, is asymmetric and acyclic;%\footnote{{ }Formally, $x$ is \textsl{revealed-to-follow} $y$ if for some $A\in\X$, either (i) $x=c(A\cup y)$ and $[y=c(\{x,y\})$ or $x\neq c(A)]$, or (ii) $x\neq c(A\cup y)$ and $[x=c(\{x,y\})$ or $x=c(A)]$.}

	\item[$\bullet$ \textsf{Rationalizable by game trees (RGT)} \citep{XuZhou2007}:] $c$ is RGT iff both \textsl{weak separability} (WS) and \textsl{divergence consistency} (DC) hold.\vs\vs
	\begin{description}
		\item[\textsf{WS}:] For any menu $A\in\X$ of size at least two, there is a partition $\{B,D\}$ of $A$ such that $c(S\cup T)=c(\{c(S),c(T)\})$ for any $S\subseteq B$ and $T\subseteq D$.\vs
		\item[\textsf{DC}:] For any $x,y,z \in X$, let $x \circlearrowleft \{y,z\}$ denote the following: $c(\{x,y,z\}) = x$, and either (i) $c(\{x,y\}) = x$, $c(\{y,z\}) = y$ and $c(\{x,z\}) = z$, or (ii) $c(\{x,y\}) = y$, $c(\{y,z\}) = z$ and $c(\{x,z\}) = x$. Then DC says that for any $x_1,x_2,y_1,y_2\in X$, if $x_1 \circlearrowleft \{y_1, y_2\}$ and $y_1 \circlearrowleft \{x_1,x_2\}$, then $c(\{x_1,y_1\}) = x_1 \:\Longleftrightarrow\: c(\{x_2,y_2\}) = y_2$. 
	\end{description}

	\item[$\bullet$ \textsf{Rational shortlist method (RSM)} \citep{ManziniMariotti2007}:] $c$ is RSM iff both \textsl{Weak WARP} (WWARP) and \textsl{property} $\gamma$ hold.\vs\vs
	\begin{description}
		\item[\textsf{WWARP}:] see below.\vs
		\item[\textsf{Property} $\gamma$:] if $c(A) = c(B)= x$, then $c(A \cup B) =x$.\vs\vs
	\end{description}
	RSM is equivalent to being \textsl{rationalizable by a post-dominance rationality procedure} \citep{RubinsteinSalant2008}, which is in turn characterized by the property of \textsl{exclusion consistency} (EC).\vs\vs
	\begin{description}
      \item[\textsf{EC}:]For any $A\in\X$ and $x\in X\setminus A$, if $c(A\cup \{x\})\notin \{c(A),x\}$, then there is no $A' \in \X$ such that $x \in A'$ and $c(A')=c(A)$.
	\end{description}

	\item[$\bullet$ \textsf{Sequentially rationalizable (SR)} \citep{ManziniMariotti2007}:] By definition, $c$ is SR iff if there is an ordered list $\L=(\succ^{1}, \ldots,\succ^{n})$ of asymmetric relations on $X$ such that for each $ A\in\X$, upon defining recursively $M_{0}(A):=A$ and $M_{i}(A):=\max(M_{i-1}(A),\succ^{i})$ for $i=1,...,n$, the equality $c(A)=M_{n}(A)$ holds.

	\item[$\bullet$ \textsf{Choice by lexicographic semiorders (CLS)} \citep{ManziniMariotti2012a}:] CLS is equivalent to being SR by an ordered list $\L=(\succ^{1}, \ldots,\succ^{n})$ of acyclic relations.
	% such that for each $ A\in\X$, upon defining recursively $M_{0}(A):=A$ and $M_{i}(A):=\max(M_{i-1}(A),\succ^{i})$ for $i=1,...,n$, the equality $c(A)=M_{n}(A)$ holds.
 %reducibility (Re) holds, where\vs
	%\begin{description}
	%	\item[\textsf{Re}:]for any $\es \neq \mathscr{S} \subseteq \X$, there is $S \in \mathscr{S}$ and $x,y \in S$ such that, for all $T \in \mathscr{S}$, if $T \setminus \{y\} \in \mathscr{S}$, then $c(T) = c ( T \setminus \{y\})$. \vs\vs
	%\end{description}
	
	\item[$\bullet$ \textsf{Weak WARP (WWARP)} \citep{ManziniMariotti2007}:] $c$ satisfies WWARP iff for any distinct $x,y \in A \subseteq B$,  $c(\{x,y\}) = c(B) = x$ implies $c(A) \neq y$. It turns out that WWARP characterizes three models of bounded rationality present in the literature, namely \textsl{categorize-then-choose} \citep{ManziniMariotti2012b}, \textsl{consistency with basic rationalization theory} \citep{CherepanovFeddersenSandroni2013}, and \textsl{overwhelming choice} \citep{LlerasMasatliogluNakajimaOzbay2017}.
	
	\item[$\bullet$ \textsf{Choice with limited attention (CLA)} \citep{MasatliogluNakajimaOzbay2012}:] $c$ is CLA iff %the relation $P$, defined by $xPy$ if there is $A\in\X$ such that $c(A)=x\neq A\setminus\{y\}$, is asymmetric and acyclic.
	\textsl{WARP with limited attention} (WARP(LA)) holds.\vs\vs 
	\begin{description}\item[\textsf{WARP(LA)}:]
	for any $A\in\X$, there is $x\in A$ such that for any $B$ containing $x$, if $c(B)\in A$ and $c(B)\neq c(B\setminus \{x\})$, then $c(B)=x$.
	\end{description}
\end{description}

Here we prove the following result:

\begin{theorem}[\cite{GiarlottaPetraliaWatson2021}, Lemma~8]\label{THM:estimates}
Let $\mathscr P$ be any of the properties (models) SQB, RGT, RSM, SR, CLS, LR, WWARP, and CLA. 
The number $q$ of non-isomorphic\footnote{{ }Two choices $c,c' \colon \X \to X$ are \textit{isomorphic} if there is a bijection $\sigma \colon X \to X$ such that $\sigma(c(A)) = c'(\sigma(A))$ for any $A \in \X$. This definition extends to choices defined on different ground sets in the obvious way.} choices on $4$ items satisfying $\mathscr P$ is
\begin{table}[H]
\small
\begin{center} 
\begin{tabular}{|c|||c|c|c|c|c|c|c|c|c|c|}
\hline
$\mathscr P$ & \;\;\footnotesize SQB\;\;& \footnotesize \;\;LR & \footnotesize \;\;RGT\;\; & \footnotesize \;\;RSM\;\; &\footnotesize \;\:\;\;SR\;\;\;\: & \;\;\; \footnotesize CLS\;\;\; & \footnotesize \!WWARP\! & \footnotesize \;\;\;CLA\;\;\; \\ 
\hline\hline
$q$ & \footnotesize $6$ & $10$  &$11$ & \footnotesize $11$ & \footnotesize $15$& \footnotesize $15$ & \footnotesize $304$ & \footnotesize $324$ \\
\hline
\end{tabular}     
\end{center}
\end{table}
\end{theorem}\vs\vs

%%%%%%%%%%%%%%%%%%%%%%%%%%%%%%%%%%%
%%%%%%%%%%%%%%%%%%%%%%%%%%%%%%%%%%%
%%%%%%%%%%%%%%%%%%%%%%%%%%%%%%%%%%%
%%%%%%%%%%%%%%%%%%%%%%%%%%%%%%%%%%%

\section{\textbf{Method summary}}

We count the number of non-isomorphic choices $c \colon \X \to X$ on $X = \{a,b,d,e\}$ satisfying any of the eight properties (models) mentioned in Theorem~\ref{THM:estimates}. 
To simplify notation, we eliminate set delimiters and commas in menus, writing $abd$ in place of $\{a,b,d\}$, $c(abd)$ in place of $c(\{a,b,d\})$, etc.
(Thus, $X = abde$.)   

For any property $\mathscr{P}$, first we derive suitable constraints from the satisfaction of $\mathscr{P}$, and then compute the number of choices satisfying these restrictions.
Note that we shall not analyze all models in Theorem~\ref{THM:estimates} in the same order as they are listed in it, but according to convenience, because some properties imply others (for instance, we have LR $\Longrightarrow$ RGT, RSM $\Longrightarrow$ SR, CLS $\Longrightarrow$ SR, and SQB $\Longrightarrow$ SR). 
To start, we make an overall computation.  

\begin{lemma} \label{LEM:all}
	The total number of non-isomorphic choices on $X$ is $864$. 
\end{lemma}	

\begin{proof}
	The problem is equivalent to counting the number of choices such that $c(abde)=a$, $c(bde)=b$, and $c(de)=d$. 
	There are $3^{{4 \choose 3}-1} \,2^{{4 \choose 2} -1} = 864$ such choices.\footnote{{ }Compare this proof with the one presented in \citet[Corollary~2]{GiarlottaPetraliaWatson2021}.} 
\end{proof}

Next, we describe the two approaches that we shall employ for all computations. 
\medskip

\noindent \textsc{\large \textbf{\underline{Approach \#1}:}}
\smallskip

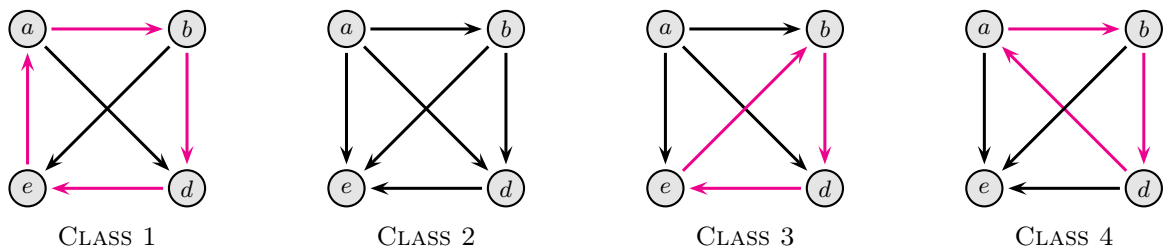
\begin{figure}[H]
\begin{center}
\psset{xunit=2.12} \psset{yunit=2.12}  
\begin{pspicture}[showgrid=false](0,-0.25)(7,1.1)   
%
%%%%%%%%%%%%%%%%%%%%%%% CLASS 1
%
\psellipse[fillstyle=solid,fillcolor=lightgray,opacity=0.4](0,0)(0.12,0.12)
\psellipse[fillstyle=solid,fillcolor=lightgray,opacity=0.4](0,1)(0.12,0.12)   
\psellipse[fillstyle=solid,fillcolor=lightgray,opacity=0.4](1,0)(0.12,0.12)
\psellipse[fillstyle=solid,fillcolor=lightgray,opacity=0.4](1,1)(0.12,0.12)
\rput(0,1){\footnotesize $a$} 
\rput(1,1){\footnotesize $b$}
\rput(1,0){\footnotesize $d$}
\rput(0,0){\footnotesize $e$}
\psline[arrowsize=5pt,linewidth=0.04,linecolor=magenta]{>}(0.85,1)(0.15,1) % a-> b
\psline[arrowsize=5pt,linewidth=0.04,linecolor=magenta]{>}(1,0.15)(1,0.85) % b -> d
\psline[arrowsize=5pt,linewidth=0.04,linecolor=magenta]{>}(0.15,0)(0.85,0) % d -> e
\psline[arrowsize=5pt,linewidth=0.04,linecolor=magenta]{>}(0,0.85)(0,0.15) % e -> a
\psline[arrowsize=5pt,linewidth=0.04]{>}(0.89,0.11)(0.11,0.89) % a -> d
\psline[arrowsize=5pt,linewidth=0.04]{>}(0.11,0.11)(0.89,0.89) % b -> e
\rput(0.5,-0.3){\small \textsc{Class 1}}
%
%%%%%%%%%%%%%%%%%%%%%%% CLASS 2
%
\psellipse[fillstyle=solid,fillcolor=lightgray,opacity=0.4](2,0)(0.12,0.12)
\psellipse[fillstyle=solid,fillcolor=lightgray,opacity=0.4](2,1)(0.12,0.12)   
\psellipse[fillstyle=solid,fillcolor=lightgray,opacity=0.4](3,0)(0.12,0.12)
\psellipse[fillstyle=solid,fillcolor=lightgray,opacity=0.4](3,1)(0.12,0.12)
\rput(2,1){\footnotesize $a$} 
\rput(3,1){\footnotesize $b$}
\rput(3,0){\footnotesize $d$}
\rput(2,0){\footnotesize $e$}
\psline[arrowsize=5pt,linewidth=0.04]{>}(2.85,1)(2.15,1) % a-> b
\psline[arrowsize=5pt,linewidth=0.04]{>}(3,0.15)(3,0.85) % b -> d
\psline[arrowsize=5pt,linewidth=0.04]{>}(2.15,0)(2.85,0) % d -> e
\psline[arrowsize=5pt,linewidth=0.04]{>}(2,0.15)(2,0.85) % a -> e
\psline[arrowsize=5pt,linewidth=0.04]{>}(2.89,0.11)(2.11,0.89) % a -> d
\psline[arrowsize=5pt,linewidth=0.04]{>}(2.11,0.11)(2.89,0.89) % b -> e
\rput(2.5,-0.3){\small \textsc{Class 2}}
%
%%%%%%%%%%%%%%%%%%%%%%%  CLASS 3
%
\psellipse[fillstyle=solid,fillcolor=lightgray,opacity=0.4](4,0)(0.12,0.12)
\psellipse[fillstyle=solid,fillcolor=lightgray,opacity=0.4](4,1)(0.12,0.12)   
\psellipse[fillstyle=solid,fillcolor=lightgray,opacity=0.4](5,0)(0.12,0.12)
\psellipse[fillstyle=solid,fillcolor=lightgray,opacity=0.4](5,1)(0.12,0.12)
\rput(4,1){\footnotesize $a$} 
\rput(5,1){\footnotesize $b$}
\rput(5,0){\footnotesize $d$}
\rput(4,0){\footnotesize $e$}
\psline[arrowsize=5pt,linewidth=0.04]{>}(4.85,1)(4.15,1) % a-> b
\psline[arrowsize=5pt,linewidth=0.04,linecolor=magenta]{>}(5,0.15)(5,0.85) % b -> d
\psline[arrowsize=5pt,linewidth=0.04,linecolor=magenta]{>}(4.15,0)(4.85,0) % d -> e
\psline[arrowsize=5pt,linewidth=0.04]{>}(4,0.15)(4,0.85) % a -> e
\psline[arrowsize=5pt,linewidth=0.04]{>}(4.89,0.11)(4.11,0.89) % a -> d
\psline[arrowsize=5pt,linewidth=0.04,linecolor=magenta]{>}(4.89,0.89)(4.11,0.11) % e -> b
\rput(4.5,-0.3){\small \textsc{Class 3}}
%
%%%%%%%%%%%%%%%%%%%%%%%%% CLASS 4
%
\psellipse[fillstyle=solid,fillcolor=lightgray,opacity=0.4](6,0)(0.12,0.12)
\psellipse[fillstyle=solid,fillcolor=lightgray,opacity=0.4](6,1)(0.12,0.12)   
\psellipse[fillstyle=solid,fillcolor=lightgray,opacity=0.4](7,0)(0.12,0.12)
\psellipse[fillstyle=solid,fillcolor=lightgray,opacity=0.4](7,1)(0.12,0.12)
\rput(6,1){\footnotesize $a$} 
\rput(7,1){\footnotesize $b$}
\rput(7,0){\footnotesize $d$}
\rput(6,0){\footnotesize $e$}
\psline[arrowsize=5pt,linewidth=0.04,linecolor=magenta]{>}(6.85,1)(6.15,1) % a-> b
\psline[arrowsize=5pt,linewidth=0.04,linecolor=magenta]{>}(7,0.15)(7,0.85) % b -> d
\psline[arrowsize=5pt,linewidth=0.04]{>}(6.15,0)(6.85,0) % d -> e
\psline[arrowsize=5pt,linewidth=0.04]{>}(6,0.15)(6,0.85) % a -> e
\psline[arrowsize=5pt,linewidth=0.04,linecolor=magenta]{>}(6.11,0.89)(6.89,0.11) % d -> a
\psline[arrowsize=5pt,linewidth=0.04]{>}(6.11,0.11)(6.89,0.89) % b -> e
\rput(6.5,-0.3){\small \textsc{Class 4}}
\end{pspicture}
\end{center}
\caption{\small The four classes in Approach\:\#1.\label{FIG:four_classes}}   
\end{figure}

We describe a graph-theoretic partition of all non-isomorphic choices on $X=abde$.
The four classes of the partition are obtained by considering all non-isomorphic selections over pairs of elements, that is, each class is associated to a digraph (see Figure~\ref{FIG:four_classes}).\vs\vs

\begin{description}
	\item[\textsc{Class} 1 (4-cycle):] $c(ab)=a$, $c(ad)=a$, $c(ae)=e$, $c(bd)=b$, $c(be)=b$, $c(de)=d$.
	In this case, the four selections $c(ab)=a$, $c(bd)=b$, $c(de)=d$, and $c(ae)=e$ reveal a cyclic binary choice, which involves all items in $X$ (in magenta in Figure~\ref{FIG:four_classes}).\vs\vs
	\item[\textsc{Class} 2 (source and sink):]$c(ab)=a$, $c(ad)=a$, $c(ae)=a$, $c(bd)=b$, $c(be)=b$, $c(de)=d$.
	In this case, the item $a$ is a \textsl{source} (because it is always selected in any binary comparison), whereas $e$ is a \textsl{sink} (because it is never chosen at a binary level).
	Note that there is no ciclic binary selection involving all four items.
	Observe also that the associated digraph is acyclic, in fact it represents the linear order $a \rhd b \rhd d \rhd e$.\vs\vs
	\item[\textsc{Class} 3 (source but no sink):] $c(ab)=a$, $c(ad)=a$, $c(ae)=a$, $c(bd)=b$, $c(be)=e$, $c(de)=d$. 
	Again, $a$ is a source, but there is no sink.
	Moreover, there is no $4$-cycle, whereas the three items different from the source create a $3$-cycle (in magenta).\vs\vs
	\item[\textsc{Class} 4 (sink but no source):] $c(ab)=a$, $c(ad)=d$, $c(ae)=a$, $c(bd)=b$, $c(be)=b$, $c(de)=d$.
	Here $e$ is a sink, but there is no source.
	Dually to Class 3, there is no 4-cycle, whereas the three items different from the sink create a $3$-cycle (in magenta). 
\end{description}

%%%%%%%%%%%%%  4 classes  %%%%%%%%%%%%%%%%%%
%
%
%%%%%%%%%%%%%%%%%%%%%%%%%%%%%%%%%

\noindent The above classes are mutually exclusive, and choices belonging to different classes are pairwise non-isomorphic.
Furthermore, any choice on $X$ is isomorphic to a choice belonging to one of these four classes. 
We conclude that Classes 1-4 provide a partition of the set of all choices to be analyzed. 
This graph-theoretic approach will be employed to count choices that are RGT, LR, SR, SQB, RSM, and CLS.
To that end, it suffices to establish the selection on 5 menus only, namely the four triples and the ground set. 
We shall do that by determining some conditions that are necessary for the model to hold. 
Then, for each choice under examination, we show that either these conditions are also sufficient, or the given model cannot satisfy them.  

Observe that this approach applies to all models of bounded rationality, as long as their definition or the behavioral properties characterizing them allow one to make enough deductions (that is, starting from the selection over pairs of items, we can determine the selection over larger menus). 
Note also that this approach naturally extends to computing the number of non-isomorphic choices on $n \geqslant 4$ items; however, as $n$ grows, this requires considering several cases, due to the large number of non-isomorphic digraphs on $n$ nodes. 
\bigskip

\noindent \textsc{\large \textbf{\underline{Approach\:\#2}:}}
\smallskip

For the remaining two models (WWARP and CLA),  we shall assume, without loss of generality, that $c$ satisfies the following conditions (see the proof of Lemma~\ref{LEM:all}): \vs\vs 
\begin{equation}\label{EQ:initial_selection}
	c(abde)=a, \quad c(bde)=b, \quad c(de)=d.\vs\vs	
\end{equation}
In this case, it suffices to determine the selection on $8$ menus, namely $4-1=3$ triples and $6-1=5$ pairs of items.
To that end, we deal with WWARP and CLA in a different way: in fact, for RSM and WWARP we provide a proof-by-cases, whereas CLA is handled by describing the code of two \textsc{Matlab} programs. 

As for Approach\:\#1, also Approach\:\#2 can be adapted to any model of bounded rationality.
Moreover, this methodology also applies to computing the number of non-isomorphic choices on $n \geqslant 4$ items (by fixing the selection over suitable $n-1$ menus). 

%%%%%%%%%%%%%%%%%%%%%%%%%%%%%%%%%%%
%%%%%%%%%%%%%%%%%%%%%%%%%%%%%%%%%%%
%%%%%%%%%%%%%%%%%%%%%%%%%%%%%%%%%%%
%%%%%%%%%%%%%%%%%%%%%%%%%%%%%%%%%%%

\section{\textbf{Method details}}

%%%%%%%%%%%%%%%%%%%%%%%%%%%%%%%%%%%
%%%%%%%%%%%%%%%%%%%%%%%%%%%%%%%%%%%

\subsection{Rationalizable by game trees (RGT)}

{\large \begin{lemma} \label{LEM:RGT}
	There are exactly $11$ non-isomorphic RGT choices on $X$.
\end{lemma}	}
 
 \begin{proof}
\cite{ApesteguiaBallester2013} show that RGT implies SR.
On the other hand, \cite{ManziniMariotti2007} prove that any SR choice satisfies \textsl{Always Chosen (AC)}:\vs\vs
\begin{description}\item[\quad \textsf{AC}:]
%	for any $A\in\X$, there is $x\in A$ such that for any $B$ containing $x$, if $c(B)\in A$ and $c(B)\neq c(B\setminus \{x\})$, then $c(B)=x$.\vs
	for any $A\in\X$ and $x \in A$, if $c(xy) = x$ for all $y \in A \setminus x$, then $c(A) = x$.\vs
	\end{description}
Thus, in particular, any RGT choice satisfies AC.
We now proceed to a proof-by-cases, distinguishing the four classes described in Approach \#1.\vs

\begin{description}
	\item[\textsc{Class} 1: (4-cycle):]$c(ab)=a$, $c(ad)=a$, $c(ae)=e$, $c(bd)=b$, $c(be)=b$, and $c(de)=d$.
	Assume $c$ is RGT, that is, WS and DC hold.
	AC implies that $c(abd)=a$, and $c(bde)=b$.
	We do not know $c(abe)$, $c(ade)$, and $c(abde)$.
	Using the definition of WS, we shall consider seven subclasses of Class $1$, which are based on all possible partitions of $X=abde$, and derive what the definition of $c$ on the three remaining menus must be. 
	Upon checking that these choices satisfy both WS and DC (and are different from each other), we obtain all possible RGT choices on $X$.\vs\vs
	\begin{description}
		\item[1A:] $abde=a \cup bde$.
		In what follows, we first make some deductions from the fact that $c$ must satisfy WS, and then derive that there is a unique choice of this kind. 
		Upon checking that WS and DC hold for $c$, we conclude that $c$ is RGT. 
		By WS, we have $c(S\cup T)=c(c(S)c(T))$ for any $S\subseteq a$ and $T\subseteq bde$.
		From $c(bde)=b$ and $c(ab)=a$, we deduce $c(abde)=a$.
		From $de \subseteq bde$, $c(de)=d$, and $c(ad)=a$, we deduce $c(ade)=a$.  
		Moreover, from $be\subseteq bde$, $c(be)=b$, and $c(ab)=a$, we deduce $c(abe)=a$.
		The reader can check that $c$ satisfies WS and DC, hence it is RGT.
		(1 RGT choice.)\vs
%Note that $c$ also satisfy DC, thus it is RGT.
%
		\item[1B:] $abde=ade\cup b$.
		By WS, $c(S\cup T)=c(c(S)c(T))$ for any $S\subseteq ade$ and $T\subseteq b$.
		Since $ae\subseteq ade$, $c(ae)=e$, and $c(be)=b$, we must have $c(abe)=b$. 
		We are still missing $c(ade)$ and $c(abde)$. 
		We distinguish three additional subcases.\vs
		\begin{description}
			\item[1Bi:] $c(ade)=a$. Since $c(ab)=a$, WS yields $c(abde)=a$.
			\item[1Bii:] $c(ade)=d$. Since $c(bd)=b$, WS yields $c(abde)=b$.
			\item[1Biii:]$c(ade)=e$. Since $c(be)=b$, WS yields $c(abde)=b$.\vs
		\end{description}
		In all subcases 1Bi, 1Bii, and 1Biii, one can check that $c$ satisfies WS and DC, hence it is RGT.
		(3 RGT choices.)\vs
		\item[1C:] $abde= abe\,\cup\,d$.
		By WS, $c(S\cup T)=c(c(S)c(T))$ for any $S\subseteq abe$ and $T\subseteq d$.
		Since $ae\subseteq abe$, and $c(ae)=e$, and $c(de)=d$, we get $c(ade)=d$.
		Again, three subcases are possible.\vs
		\begin{description}
			\item[1Ci:]$c(abe)=a$. Since $c(ad)=a$, WS yields $c(abde)=a$.
			\item[1Cii:]$c(abe)=b$.  Since $c(bd)=b$, WS yields $c(abde)=b$.
			\item[1Ciii:]$c(abe)=e$. Since $c(de)=d$, WS yields $c(abde)=d$.\vs
		\end{description}
		In all subcases 1Ci, 1Cii, and 1Ciii, $c$ satisfies WS and DC, hence it is RGT.
		(3 RGT choices.)\vs
		\item[1D:] $abde=abd\cup e$.
		WS yields $c(S\cup T)=c(c(S)c(T))$ for any $S\subseteq abd$ and $T\subseteq e$.
		Since $ab\subseteq abd$, $c(ab)=a$, and $c(ae)=e$, we get $c(abe)=e$.
		Since $ad\subseteq abd$, $c(ad)=a$, and $c(ae)=e$,  we get $c(ade)=e$.
		Finally, since $c(abd)=a$, and $c(ae)=e$, we get $c(abde)=e$.
		This choice $c$ satisfies WS and DC, hence it is RGT.
		(1 RGT choice.)\vs
		\item[1E:] $abde=ab\cup de$.
		WS yields $c(S\cup T)=c(c(S)c(T))$ for any $S\subseteq ab$ and $T\subseteq de$.
		From $c(ab)=a$, $c(de)=d$, and $c(ad)=a$, we deduce $c(abde)=a$.
		From $e\subseteq de$, $c(ab)=a$, and $c(ae)=e$, we deduce $c(abe)=e$.
		From $a\subseteq ab$, $c(de)=d$, and $c(ad)=a$, we deduce $c(ade)=a$.
		This choice $c$ satisfies WS and DC, hence it is RGT.
		(1 RGT choice.)\vs
		\item[1F:] $abde=ad\cup be$.
		WS yields $c(S\cup T)=c(c(S)c(T))$ for any $S\subseteq ad$ and $T\subseteq be$.
		Since $c(ad)=a$, $c(be)=b$, and $c(ab)=a$, we deduce $c(abde)=a$.
		Since $a\subseteq ad$, $c(be)=b$, and $c(ab)=a$, we deduce $c(abe)=a$.
		Since $e\subseteq be$, $c(ad)=a$, and $c(ae)=e$, we deduce $c(ade)=e$.
		This choice $c$ satisfies WS.
		However, DC fails for $c$, because we have $e \circlearrowleft ad$, $a \circlearrowleft  be$, $c(ea) = e$, and yet $c(db) =b$.\footnote{{ }We are taking $x_1 := e$, $x_2:=b$, $y_1:=a$, and $y_2:=d$ in the definition of DC.}   
		It follows that $c$ is not RGT.
		(0 RGT choice.)\vs
		 \item[1G:] $abde=ae\cup bd$.
 		WS yields $c(S\cup T)=c(c(S)c(T))$ for any $S\subseteq ae$ and $T\subseteq bd$.
 		From $c(ae)=e$, $c(bd)=b$, and $c(be)=b$, we get $c(abde)=b$.
 		From $b\subseteq bd$, $c(ae)=e$, and $c(be)=b$, we get $c(abe)=b$.
 		From $d\subseteq bd$, $c(ae)=e$, and $c(de)=d$, we get $c(ade)=d$.
 		This choice $c$ satisfies WS and DC, hence it is RGT.
 		(1 RGT choice.)\vs\vs
 	\end{description}
In Class~$1$, WS does not hold for any choice different from those listed above.
Note also that choices defined in subcases 1Bii, 1Cii, and 1G are the same.
We conclude that in Class 1 there are exactly $8 =10-2$ pairwise non-isomorphic  RGT choices.\vs

 \item[\textsc{Class 2} (source and sink):]$c(ab)=a$, $c(ad)=a$, $c(ae)=a$, $c(bd)=b$, $c(be)=b$, $c(de)=d$.
   Assume $c$ is RGT.
  AC readily implies that $c(abd)=c(abe)=c(ade)=c(abde)=a$, and $c(bde)=b$.
  Thus, in this class we get a unique choice $c$, which is rationalizable, and so it is also RGT.\vs 
 
 \item[\textsc{Class} 3 (source but no sink):] $c(ab)=a$, $c(ad)=a$, $c(ae)=a$, $c(bd)=b$, $c(be)=e$, $c(de)=d$.
 Assume $c$ is RGT.
 By AC, we get $c(abd)=c(abe)=c(ade)=c(abde)=a$.
 Without loss of generality, we can assume $c(bde)=b$.\footnote{{ }Indeed, the other two subcases, namely $c(bde)=d$ and $c(bde)=e$, generate choices that are isomorphic to the one we are considering. For instance, if $c(bde)=d$, then the $3$-cycle $\langle b,d,e \rangle$, which is defined by $a \mapsto a$ and $b \mapsto d \mapsto e \mapsto b$, is a choice isomorphism from $X$ onto $X$.}
 The reader can check that $c$ satisfies WS and DC, hence it is RGT.\vs

\item[\textsc{Class} 4 (sink but no source):] $c(ab)=a$, $c(ad)=d$, $c(ae)=a$, $c(bd)=b$, $c(be)=b$, $c(de)=d$.
Assume $c$ is RGT.
By AC, we get $c(abe)=a$, $c(ade)=d$, and $c(bde)=b$.
Without loss of generality, we can assume $c(abd)=a$.\footnote{{ }As in Class~3, the other two subcases $c(abd)=b$ and $c(abd)=d$ give isomorphic choices.}  
We do not know $c(abde)$.
As for Class~1, we examine all possible partitions of $abde$ that are compatible with WS.

To start, we claim that we can discard all partitions of $X$ in which the two items $b,d$ do not belong to the same subset of $abde$.
To see why, assume by way of contradiction that $c$ satisfies WS for a partition $X_1 \cup X_2$ of $abde$ such that $b\in X_1$ and $d\in X_2$.
Note that $a$ may belong to $X_1$ or $X_2$. 
Suppose $a\in X_1$. 
Since $ab \subseteq X_1$, $d \subseteq X_2$, $c(ab)=a$, and $c(ad)=d$, WS yields $c(abd) = d$, which contradicts the hypothesis $c(abd)=a$. 
Thus, $a\in X_2$ holds.
However, since $b \subseteq X_1$, $ad \subseteq X_2$, $c(ad)=d$, and $c(bd)=b$, now WS yields $c(abd)=b$, which is again a contradiction. 
This proves the claim.

By virtue of the above claim, we may only consider partitions of the type $abde=X_1\cup X_2$ such that $b,d\in X_1$, or $b,d\in X_2$. 
Three subcases arise.\vs\vs
\begin{description}
	\item[\textsc{4A}:] $abde=ae\cup bd$. 
 	By WS, $c(S\cup T)=c(c(S)c(T))$ for any $S\subseteq ae$ and $T\subseteq bd$.
 	Since $c(ae)=a$, $c(bd)=b$, and $c(ab)=a$, we obtain $c(abde)=a$.\vs
	\item[{4B}:] $abde=abd\cup e$. 
	By WS, $c(S\cup T)=c(c(S)c(T))$ for any $S\subseteq abd$ and $T\subseteq e$.
 	Since $c(abd)=a$ and $c(ae)=a$, we obtain $c(abde)=a$.\vs
 	\item[\textsc{4C}:] $abde=a\cup bde$. 
 	By WS, $c(S\cup T)=c(c(S)c(T))$ for any $S\subseteq a$ and $T\subseteq bde$.
 	Since $c(bde)=b$ and $c(ab)=a$, we obtain $c(abde)=a$.
\end{description}
 
 Therefore 4A, 4B, and 4C all generate the same choice $c$.
 The reader can check that $c$ satisfies WS and DC.
 Overall, Class 4 only gives 1 RGT choice.\vs\vs 
\end{description}

\noindent Summing up Classes 1-4, we obtain 8+1+1+1=11 non-isomorphic RGT choices on $X$. 
\end{proof}

%%%%%%%%%%%%%%%%%%%%%%%%%%%%%%%%%%%
%%%%%%%%%%%%%%%%%%%%%%%%%%%%%%%%%%%

\subsection{List rational (LR)}

{\large \begin{lemma} \label{LEM:LR}
There are exactly $10$ non-isomorphic LR choices on $X$.
\end{lemma}}

 \begin{proof}
\cite{Yildiz2016} states that any LR choice is RGT.
%Thus, LR implies WS and DC.
In Lemma~\ref{LEM:RGT}, we have described 11 non-isomorphic RGT choices on $X$.
Below we shall show that all but one of the 11 RGT choices are LR. 
Specifically, for each of these 11 RGT choices, first we determine some obvious necessary conditions for being LR, and then we prove that these necessary conditions are either sufficient (for 10 choices) or impossible (for 1 choice).\footnote{{ }It suffices to check that the equality $c(A) = c(c(A\setminus x)x)$ holds for any menu $A$ such that $\vert A \vert \geqslant 3$.}
We use the same numeration as in the proof of Lemma~\ref{LEM:RGT}.

\begin{description}
	\item[\textsc{1A}:]$c(ab)=a$, $c(ad)=a$, $c(ae)=e$, $c(bd)=b$, $c(be)=b$, $c(de)=d$, $c(abd)=a$, $c(abe)=a$, $c(ade)=a$, $c(bde)=b$, $c(abde)=a$. 
	Assume $c$ is LR. 
	By definition, there is a linear order $\rhd$ on $X$ such that $c(A)=c(c(A\setminus{x})x)$ for any $A\in\X$, where $x = \min(A,\rhd)$.
	
	\textsc{Claim\:1:} $b\rhd a$ and $e \rhd a$.
	To prove it, we use the fact that $c(ae)=e$ and $c(abe)=a$.
	Toward a contradiction, suppose $a \rhd b$ or $a\rhd e$.
	Three cases are possible: (1) $a\rhd b$ and $e\rhd a$; (2) $b\rhd a$ and $a\rhd e$; (3) $a\rhd b$ and $a\rhd e$. 
	In case (1), transitivity of $\rhd$ yields $e \rhd b$, and so $\min(abe,\rhd) = b$. 
	By hypothesis, we obtain $c(abe)=c(c(ae)b)=c(be)=b \neq a$, a contradiction. 
	In case (2), transitivity of $\rhd$ yields $b \rhd e$, and so $\min(abe,\rhd) = e$. 
	By hypothesis, we obtain $c(abe)=c(c(ab)e)=c(ae)=e\neq a$, a contradiction. 
	In case (3), $e\rhd  b$ implies $c(abe)=c(c(ae)b)=c(be)=b\neq a$, whereas $b\rhd  e$ implies $c(abe)=c(c(ab)e)=c(ae)=e\neq a$, a contradiction in both circumstances.
	
	\textsc{Claim\:2:} $d\rhd a$ and $e\rhd a$.
	The proof of Claim\:2 is similar to that of Claim\:1, using the fact that $c(ae)=e$ and $c(ade)=a$.
	
	Summarizing, Claims\:1\:and\:2 yield the necessary conditions $b \rhd a$, $d \rhd a$, $e \rhd a$.
	Thus, the list $\rhd$ must extend the partial order\footnote{{ }Recall that a \textsl{partial order} is a reflexive, transitive, and antisymmetric binary relation.} associated to the following Hasse diagram:\footnote{{ }In a \textsl{Hasse Diagram}, a segment from $x$ (top) to $y$ (bottom) stands for $x \rhd y$, and transitivity is always assumed to hold (thus, two consecutive segments from $x$ to $y$, and from $y$ to $z$ stand for $x \rhd y$, $y \rhd z$, $x \rhd z$).}	
%%%%%%%%%%%%%  LR class 1A  %%%%%%%%%%%%%%%%%%
%
%\begin{figure}[H]
%\begin{center}
%\psset{xunit=2.12} \psset{yunit=2.12}  
%\begin{pspicture}[showgrid=false](0,0.2)(1,1.1)   
%%
%%
%\psellipse[fillstyle=solid,fillcolor=lightgray,opacity=0.4](0,0)(0.12,0.12)
%\psellipse[fillstyle=solid,fillcolor=lightgray,opacity=0.4](0,1)(0.12,0.12)   
%\psellipse[fillstyle=solid,fillcolor=lightgray,opacity=0.4](1,0)(0.12,0.12)
%\psellipse[fillstyle=solid,fillcolor=lightgray,opacity=0.4](1,1)(0.12,0.12)
%%
%\rput(0,1){\footnotesize $a$} 
%\rput(1,1){\footnotesize $b$}
%\rput(1,0){\footnotesize $d$}
%\rput(0,0){\footnotesize $e$}
%%
%\psline[arrowsize=5pt,linewidth=0.04]{>}(0.15,1)(0.85,1) % b-> a
%%\psline[arrowsize=5pt,linewidth=0.04]{>}(1,0.15)(1,0.85) % b -> d
%%\psline[arrowsize=5pt,linewidth=0.04]{>}(0.15,0)(0.85,0) % d -> e
%\psline[arrowsize=5pt,linewidth=0.04]{>}(0,0.85)(0,0.15) % e -> a
%\psline[arrowsize=5pt,linewidth=0.04]{>}(0.11,0.89)(0.89,0.11) % d -> a
%%\psline[arrowsize=5pt,linewidth=0.04]{>}(0.11,0.11)(0.89,0.89) % b -> e
%%
%%
%\end{pspicture}
%\end{center}
%%\caption{\small The four classes in Approach \#1.\label{FIG:four_classes}}   
%%
%\end{figure}
%
%%%%%%%%%%%%%%%%%%%%%%%%%%%%%%%%%
%%%%%%%%%%%%  LR Class 1A  %%%%%%%%%%%%%%%%
%
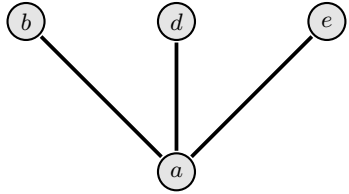
\begin{figure}[H]
\begin{center}
\psset{xunit=2} \psset{yunit=2}  
\begin{pspicture}[showgrid=false](-3,1.3)(5,2.1)  
%
%
%\psline[arrowsize=6pt,linewidth=0.05](0.85,1.85)(0.15,1.15)
%\psline[arrowsize=6pt,linewidth=0.05](1.15,1.85)(1.85,1.15)
%\psline[arrowsize=6pt,linewidth=0.05](0.83,0.17)(0.15,0.85)
%\psline[arrowsize=6pt,linewidth=0.05](1.17,0.17)(1.85,0.85)
%
\psline[arrowsize=6pt,linewidth=0.05](0.1,1.9)(0.9,1.1)
\psline[arrowsize=6pt,linewidth=0.05](1,1.86)(1,1.14)
\psline[arrowsize=6pt,linewidth=0.05](1.9,1.9)(1.1,1.1)
\psellipse[fillstyle=solid,fillcolor=lightgray,opacity=0.4](1,2)(0.13,0.13)
\psellipse[fillstyle=solid,fillcolor=lightgray,opacity=0.4](0,2)(0.13,0.13)   
\psellipse[fillstyle=solid,fillcolor=lightgray,opacity=0.4](2,2)(0.13,0.13)
\psellipse[fillstyle=solid,fillcolor=lightgray,opacity=0.4](1,1)(0.13,0.13)
\rput(1,2){\footnotesize $d$} 
\rput(0,2){\footnotesize $b$}
\rput(2,2){\footnotesize $e$}
\rput(1,1){\footnotesize $a$}
\end{pspicture}
\end{center}
\end{figure}
%
%%%%%%%%%%%%%%%%%%%%%%%%%%%%%%%%%	 
	To complete the analysis, we check that any linear order $\rhd$ extending this partial order list-rationalizes $c$. 
	It suffices to show that $c(A) = c(c(A \setminus x)x)$ for any $A \in \X$ of size at least 3, where $x = \min(A,\rhd)$. 
	Indeed, we have (regardless of how $\rhd$ ranks $b,d,e$):\vs\vs
	\begin{itemize}
		\item $c(abd) = c(c(bd)a) = c(ab) = a\,$; 
		\item $c(abe) = c(c(be)a) = c(ab) = a\,$; 
		\item $c(ade) = c(c(de)a) = c(ad) = a\,$;
		\item $c(bde) = b$ (by considering all possible cases: $\min(bde,\rhd) = e$ implies $c(bde) = c(c(bd)e)= c(be) = b$, $\min(bde,\rhd) = d$ implies $c(bde) = c(c(be)d)= c(bd) = b$, and $\min(bde,\rhd) = b$ implies $c(bde) = c(c(de)b)= c(bd) = b$); 
		\item $c(abde) = c(c(bde)a) = c(ab) = a\,$.
	\end{itemize}

	\item[1Bi:]$c(ab)=a$, $c(ad)=a$, $c(ae)=e$, $c(bd)=b$, $c(be)=b$, $c(de)=d$, $c(abd)=a$, $c(abe)=b$, $c(ade)=a$, $c(bde)=b$, $c(abde)=a$.
	(Note that this choice only differs from 1A in the selection from the menu $abe$.)  
	Assume $c$ is LR.
	Since $c(ab)=a$ and $c(abe)=b$, an argument similar to that used to prove Claim\:1 yields $a\rhd b$ and $e\rhd b$.
	Similarly, from $c(ae)=e$  and $c(ade)=a$, we derive $d\rhd a$ and $e\rhd a$.
	Thus, if $\rhd$ list-rationales $c$, then we must have $d,e \, \rhd\, a \rhd b$ (hence $d,e \rhd b$ by transitivity).
	Representing these necessary conditions by a Hasse diagram, the list $\rhd$ must extend the partial order
	%%%%%%%%%%%%%  LR Class 1B(i)  %%%%%%%%%%%%%%%%
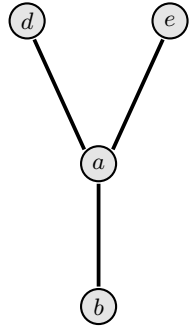
\begin{figure}[H]
\begin{center}
\psset{xunit=1.9} \psset{yunit=1.9}  
\begin{pspicture}[showgrid=false](-3,0.3)(4,2.1) 
\psline[arrowsize=6pt,linewidth=0.05](0.05,1.87)(0.4,1.1)
\psline[arrowsize=6pt,linewidth=0.05](0.95,1.87)(0.6,1.1)
\psline[arrowsize=6pt,linewidth=0.05](0.5,0.86)(0.5,0.14) 
\psellipse[fillstyle=solid,fillcolor=lightgray,opacity=0.4](1,2)(0.13,0.13)
\psellipse[fillstyle=solid,fillcolor=lightgray,opacity=0.4](0,2)(0.13,0.13)   
\psellipse[fillstyle=solid,fillcolor=lightgray,opacity=0.4](0.5,0)(0.13,0.13)
\psellipse[fillstyle=solid,fillcolor=lightgray,opacity=0.4](0.5,1)(0.13,0.13)
\rput(1,2){\footnotesize $e$} 
\rput(0,2){\footnotesize $d$}
\rput(0.5,0){\footnotesize $b$}
\rput(0.5,1){\footnotesize $a$}
\end{pspicture}
\end{center}
\end{figure}
%%%%%%%%%%%%%%%%%%%%%%%%%%%%%%%%%%	
	Now we check that these necessary conditions are also sufficient, that is, $c(A) = c(c(A \setminus x)x)$ for any $A \in \X$ of size at least 3, where $x = \min(A,\rhd)$. 
	Indeed, we have:\vs\vs
	\begin{itemize}
		\item $c(abd) = c(c(ad)b) = c(ab) = a\,$; 
		\item $c(abe) = c(c(ae)b) = c(be) = b\,$; 
		\item $c(ade) = c(c(de)a) = c(ad) = a\,$; 
		\item $c(bde) =c(c(de)b) =  c(bd) = b\,$; 
		\item $c(abde) = c(c(ade)b) = c(ab) = a\,$.
	\end{itemize}

	\item[1Bii $\equiv$ 1Cii $\equiv$ 1G:]$c(ab)=a$, $c(ad)=a$, $c(ae)=e$, $c(bd)=b$, $c(be)=b$, $c(de)=d$, $c(abd)=a$, $c(abe)=b$, $c(ade)=d$, $c(bde)=b$, $c(abde)=b$.
	Assume $c$ is LR.
	From $c(ab)=a$ and $c(abe)=b$, we derive $a\rhd b$ and $e\rhd b$.
	From $c(ad)=a$  and $c(ade)=d$, we derive $a\rhd d$ and $e\rhd d$.
	Thus, $\rhd$ must extend the partial order
	%%%%%%%%%%%%  LR Class 1B(ii)  %%%%%%%%%%%%%%%%
%
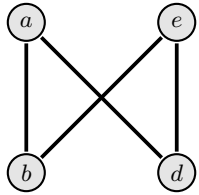
\begin{figure}[H]
\begin{center}
\psset{xunit=2} \psset{yunit=2}  
\begin{pspicture}[showgrid=false](-3,1.3)(4,2.1)  
%
%
%\psline[arrowsize=6pt,linewidth=0.05](0.85,1.85)(0.15,1.15)
%\psline[arrowsize=6pt,linewidth=0.05](1.15,1.85)(1.85,1.15)
%\psline[arrowsize=6pt,linewidth=0.05](0.83,0.17)(0.15,0.85)
%\psline[arrowsize=6pt,linewidth=0.05](1.17,0.17)(1.85,0.85)
%
\psline[arrowsize=6pt,linewidth=0.05](0.1,1.9)(0.9,1.1)
\psline[arrowsize=6pt,linewidth=0.05](1,1.86)(1,1.14)
\psline[arrowsize=6pt,linewidth=0.05](0,1.86)(0,1.14)
\psline[arrowsize=6pt,linewidth=0.05](0.9,1.9)(0.1,1.1)
\psellipse[fillstyle=solid,fillcolor=lightgray,opacity=0.4](1,2)(0.13,0.13)
\psellipse[fillstyle=solid,fillcolor=lightgray,opacity=0.4](0,2)(0.13,0.13)   
\psellipse[fillstyle=solid,fillcolor=lightgray,opacity=0.4](0,1)(0.13,0.13)
\psellipse[fillstyle=solid,fillcolor=lightgray,opacity=0.4](1,1)(0.13,0.13)
\rput(1,2){\footnotesize $e$} 
\rput(0,2){\footnotesize $a$}
\rput(0,1){\footnotesize $b$}
\rput(1,1){\footnotesize $d$}
\end{pspicture}
\end{center}
\end{figure}
%
%%%%%%%%%%%%%%%%%%%%%%%%%%%%%%%%%	 
We check that these necessary conditions are also sufficient.\vs\vs 
	\begin{itemize}
		\item $c(abd) = a\,$: If $\min(abd,\rhd)=b$, then $c(abd)=c(c(ad)b)=c(ab)=a$. Similarly, if $\min(abd,\rhd)=d$, then $c(abd)=c(c(ab)d)=c(ad)=a$.
		\item $c(abe) = c(c(ae)b) = c(be) = b$.
		\item $c(ade) = c(c(ae)d) = c(de) = d$.
		\item $c(bde) = b\,$: If $\min(bde,\rhd)=b$, then $c(bde)=c(c(de)b)=c(bd)=b$. Similarly, if $\min(bde,\rhd)=d$, then $c(bde)=c(c(be)d)=c(bd)=b$.
		\item $c(abde) = b\,$: If $\min(abde,\rhd)=b$, then $c(abde)=c(c(ade)b)=c(bd)=b$. If $\min(abde,\rhd)=d$, then $c(abde)=c(c(abe)d)=c(bd)=b$.
	\end{itemize}

	\item[1Biii:]$c(ab)=a$, $c(ad)=a$, $c(ae)=e$, $c(bd)=b$, $c(be)=b$, $c(de)=d$, $c(abd)=a$, $c(abe)=b$, $c(ade)=e$, $c(bde)=b$, $c(abde)=b$.
	Assume $c$ is LR.
	From $c(ab)=a$ and $c(abe)=b$, we get $a\rhd b$ and $e\rhd b$.
	From $c(de)=d$ and $c(ade)=e$, we get $d\rhd e$ and $a\rhd e$.
		Thus, $\rhd$ extends a partial order that is isomorphic to that of case 1Bi: 
	%%%%%%%%%%%%%  LR Class 1B(iii)  %%%%%%%%%%%%%%%%
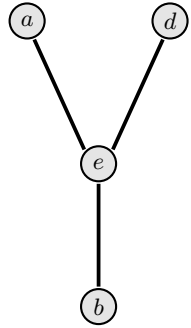
\begin{figure}[H]
\begin{center}
\psset{xunit=1.9} \psset{yunit=1.9}  
\begin{pspicture}[showgrid=false](-3,0.3)(4,2.05) 
\psline[arrowsize=6pt,linewidth=0.05](0.05,1.87)(0.4,1.1)
\psline[arrowsize=6pt,linewidth=0.05](0.95,1.87)(0.6,1.1)
\psline[arrowsize=6pt,linewidth=0.05](0.5,0.86)(0.5,0.14) 
\psellipse[fillstyle=solid,fillcolor=lightgray,opacity=0.4](1,2)(0.13,0.13)
\psellipse[fillstyle=solid,fillcolor=lightgray,opacity=0.4](0,2)(0.13,0.13)   
\psellipse[fillstyle=solid,fillcolor=lightgray,opacity=0.4](0.5,0)(0.13,0.13)
\psellipse[fillstyle=solid,fillcolor=lightgray,opacity=0.4](0.5,1)(0.13,0.13)
\rput(1,2){\footnotesize $d$} 
\rput(0,2){\footnotesize $a$}
\rput(0.5,0){\footnotesize $b$}
\rput(0.5,1){\footnotesize $e$}
\end{pspicture}
\end{center}
\end{figure}
%%%%%%%%%%%%%%%%%%%%%%%%%%%%%%%%%%	
We check that any extension of the above partial order list-rationales $c$.\vs\vs 
\begin{itemize}
		\item $c(abd) = c(c(ad)b) = c(ab) = a$. 
		\item $c(abe) = c(c(ae)b) = c(be) = b$. 
		\item $c(ade) = c(c(ad)e) = c(ae) = e$. 
		\item $c(bde) =c(c(de)b) =  c(bd) = b$. 
		\item $c(abde) = c(c(ade)b) = c(be) = b$.
	\end{itemize}

 	\item[1Ci:]$c(ab)=a$, $c(ad)=a$, $c(ae)=e$, $c(bd)=b$, $c(be)=b$, $c(de)=d$, $c(abd)=a$, $c(abe)=a$, $c(ade)=d$, $c(bde)=b$, $c(abde)=a$.
 	Assume $c$ is LR.
 	From $c(ae)=e$ and $c(abe)=a$, we derive $e\rhd a$ and $b\rhd a$.
 	From $c(ad)=a$ and $c(ade)=d$, we derive $a\rhd d$ and $e\rhd d$. Thus, $\rhd$ extends a partial order isomorphic to 1Bi and 1Bii: 
 		%%%%%%%%%%%%%  LR Class 1C(i)  %%%%%%%%%%%%%%%%
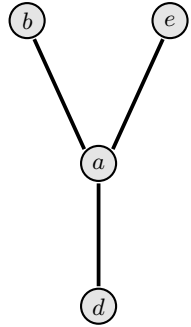
\begin{figure}[H]
\begin{center}
\psset{xunit=1.9} \psset{yunit=1.9}  
\begin{pspicture}[showgrid=false](-3,0.3)(4,2.05) 
\psline[arrowsize=6pt,linewidth=0.05](0.05,1.87)(0.4,1.1)
\psline[arrowsize=6pt,linewidth=0.05](0.95,1.87)(0.6,1.1)
\psline[arrowsize=6pt,linewidth=0.05](0.5,0.86)(0.5,0.14) 
\psellipse[fillstyle=solid,fillcolor=lightgray,opacity=0.4](1,2)(0.13,0.13)
\psellipse[fillstyle=solid,fillcolor=lightgray,opacity=0.4](0,2)(0.13,0.13)   
\psellipse[fillstyle=solid,fillcolor=lightgray,opacity=0.4](0.5,0)(0.13,0.13)
\psellipse[fillstyle=solid,fillcolor=lightgray,opacity=0.4](0.5,1)(0.13,0.13)
\rput(1,2){\footnotesize $e$} 
\rput(0,2){\footnotesize $b$}
\rput(0.5,0){\footnotesize $d$}
\rput(0.5,1){\footnotesize $a$}
\end{pspicture}
\end{center}
\end{figure}
%%%%%%%%%%%%%%%%%%%%%%%%%%%%%%%%%%	

We check that any extension of this partial order list-rationales $c$.\vs\vs
\begin{itemize}
		\item $c(abd) = c(c(ab)d) = c(ad) = a$. 
		\item $c(abe) = c(c(be)a) = c(ab) = a$.
		\item $c(ade) = c(c(ae)d) = c(de) = d$. 
		\item $c(bde) =c(c(be)d) =  c(bd) = b$. 
		\item $c(abde) = c(c(abe)d) = c(ad) = a$.
	\end{itemize}

	\item[1Ciii:]$c(ab)=a$, $c(ad)=a$, $c(ae)=e$, $c(bd)=b$, $c(be)=b$, $c(de)=d$, $c(abd)=a$, $c(abe)=e$, $c(ade)=d$, $c(bde)=b$, $c(abde)=d$.
	Assume $c$ is LR.
	From $c(be)=b$ and $c(abe)=e$, we get $a\rhd e$ and $b\rhd e$.
	From $c(ad)=a$ and $c(ade)=d$, we get $a\rhd d$ and $e\rhd d$.
	Thus, $\rhd$ extends a partial order isomorphic to the one in 1Bi, 1Bii, and 1Ci:
	 		%%%%%%%%%%%%%  LR Class 1C(iii)  %%%%%%%%%%%%%%%%
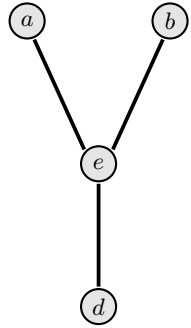
\begin{figure}[H]
\begin{center}
\psset{xunit=1.9} \psset{yunit=1.9}  
\begin{pspicture}[showgrid=false](-3,0.3)(4,2.05) 
\psline[arrowsize=6pt,linewidth=0.05](0.05,1.87)(0.4,1.1)
\psline[arrowsize=6pt,linewidth=0.05](0.95,1.87)(0.6,1.1)
\psline[arrowsize=6pt,linewidth=0.05](0.5,0.86)(0.5,0.14) 
\psellipse[fillstyle=solid,fillcolor=lightgray,opacity=0.4](1,2)(0.13,0.13)
\psellipse[fillstyle=solid,fillcolor=lightgray,opacity=0.4](0,2)(0.13,0.13)   
\psellipse[fillstyle=solid,fillcolor=lightgray,opacity=0.4](0.5,0)(0.13,0.13)
\psellipse[fillstyle=solid,fillcolor=lightgray,opacity=0.4](0.5,1)(0.13,0.13)
\rput(1,2){\footnotesize $b$} 
\rput(0,2){\footnotesize $a$}
\rput(0.5,0){\footnotesize $d$}
\rput(0.5,1){\footnotesize $e$}
\end{pspicture}
\end{center}
\end{figure}
%%%%%%%%%%%%%%%%%%%%%%%%%%%%%%%%%%	
We check that any extension of this partial order list-rationales $c$.\vs\vs
\begin{itemize}
		\item $c(abd) = c(c(ab)d) = c(ad) = a$. 
		\item $c(abe) = c(c(ab)e) = c(ae) = e$. 
		\item $c(ade) = c(c(ae)d) = c(de) = d$. 
		\item $c(bde) =c(c(be)d) =  c(bd) = b$. 
		\item $c(abde) = c(c(abe)d) = c(de) = d$.
	\end{itemize}

	\item[1D:]$c(ab)=a$, $c(ad)=a$, $c(ae)=e$, $c(bd)=b$, $c(be)=b$, $c(de)=d$, $c(abd)=a$, $c(abe)=e$, $c(ade)=e$, $c(bde)=b$, $c(abde)=e$.
	Assume $c$ is LR.
	From $c(be)=b$ and $c(abe)=e$, we get $b\rhd e$ and $a\rhd e$.
	From $c(de)=d$ and $c(ade)=e$, we get $d\rhd e$ and $a\rhd e$.
	Thus, $\rhd$ extends a partial order that is isomorphic to 1A:
	%%%%%%%%%%%%  LR Class 1D  %%%%%%%%%%%%%%%%
%
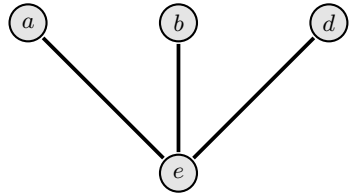
\begin{figure}[H]
\begin{center}
\psset{xunit=2} \psset{yunit=2}  
\begin{pspicture}[showgrid=false](-3,1.3)(5,2.1)  
%
%
%\psline[arrowsize=6pt,linewidth=0.05](0.85,1.85)(0.15,1.15)
%\psline[arrowsize=6pt,linewidth=0.05](1.15,1.85)(1.85,1.15)
%\psline[arrowsize=6pt,linewidth=0.05](0.83,0.17)(0.15,0.85)
%\psline[arrowsize=6pt,linewidth=0.05](1.17,0.17)(1.85,0.85)
%
\psline[arrowsize=6pt,linewidth=0.05](0.1,1.9)(0.9,1.1)
\psline[arrowsize=6pt,linewidth=0.05](1,1.86)(1,1.14)
\psline[arrowsize=6pt,linewidth=0.05](1.9,1.9)(1.1,1.1)
\psellipse[fillstyle=solid,fillcolor=lightgray,opacity=0.4](1,2)(0.13,0.13)
\psellipse[fillstyle=solid,fillcolor=lightgray,opacity=0.4](0,2)(0.13,0.13)   
\psellipse[fillstyle=solid,fillcolor=lightgray,opacity=0.4](2,2)(0.13,0.13)
\psellipse[fillstyle=solid,fillcolor=lightgray,opacity=0.4](1,1)(0.13,0.13)
\rput(1,2){\footnotesize $b$} 
\rput(0,2){\footnotesize $a$}
\rput(2,2){\footnotesize $d$}
\rput(1,1){\footnotesize $e$}
\end{pspicture}
\end{center}
\end{figure}
%
%%%%%%%%%%%%%%%%%%%%%%%%%%%%%%%%%	 
	We check that any extension of $\rhd$ list-rationalizes $c$.\vs\vs
	\begin{itemize}
		\item $c(abd) = a\,$: If $\min(abd,\rhd) = a$, then $c(abd) = c(c(bd)a)= c(ab) = a$. If $\min(abd,\rhd) = b$, then $c(abd) = c(c(ad)b)= c(ab) = a$. If $\min(abd,\rhd) = d$, then $c(abd) = c(c(ab)d)= c(ad) = a$. 
		\item $c(abe) = c(c(ab)e) = c(ae) = e$. 
		\item $c(ade) = c(c(ad)e) = c(ae) = e$.
		\item $c(bde) = c(c(bd)e) = c(be) = b$. 
		\item $c(abde) = c(c(abd)e) = c(ae) = e$.
	\end{itemize}

	\item[1E:]$c(ab)=a$, $c(ad)=a$, $c(ae)=e$, $c(bd)=b$, $c(be)=b$, $c(de)=d$, $c(abd)=a$, $c(abe)=e$, $c(ade)=a$, $c(bde)=b$, $c(abde)=a$.
 	Assume $c$ is LR.
	From $c(be)=b$ and $c(abe)=e$, we obtain $b\rhd e$ and $a\rhd e$.
	From $c(ae)=e$ and $c(ade)=a$, we obtain $e\rhd a$ and $d\rhd a$. 
	It follows that $a\rhd e\rhd a$, which is impossible.	
	We conclude that $c$ is \textbf{not} LR.\vs

	\item[2:]$c(ab)=a$, $c(ad)=a$, $c(ae)=a$, $c(bd)=b$, $c(be)=b$, $c(de)=d$, $c(abd)=a$, $c(abe)=a$, $c(ade)=a$, $c(bde)=b$, $c(abde)=a$.
	This choice is rationalizable, hence it is LR.\vs
	
	\item[3:]$c(ab)=a$, $c(ad)=a$, $c(ae)=a$, $c(bd)=b$, $c(be)=e$, $c(de)=d$, $c(abd)=a$, $c(abe)=a$, $c(ade)=a$, $c(bde)=b$, $c(abde)=a$.
	Assume $c$ is LR.
	From $c(be)=e$ and $c(bde)=b$, we derive $e\rhd b$ and $d\rhd b$.
	Thus, $\rhd$ must extend the following partial order: 
		%%%%%%%%%%%%  LR Class 3  %%%%%%%%%%%%%%%%
%
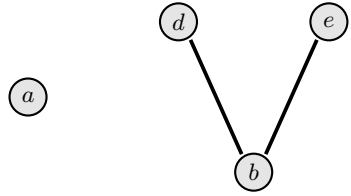
\begin{figure}[H]
\begin{center}
\psset{xunit=2} \psset{yunit=2}  
\begin{pspicture}[showgrid=false](-3,1.3)(5,2.1)  
\psline[arrowsize=6pt,linewidth=0.05](1.08,1.88)(1.42,1.12)
\psline[arrowsize=6pt,linewidth=0.05](1.92,1.88)(1.58,1.12)
\psellipse[fillstyle=solid,fillcolor=lightgray,opacity=0.4](1,2)(0.13,0.13)
\psellipse[fillstyle=solid,fillcolor=lightgray,opacity=0.4](0,1.5)(0.13,0.13)   
\psellipse[fillstyle=solid,fillcolor=lightgray,opacity=0.4](2,2)(0.13,0.13)
\psellipse[fillstyle=solid,fillcolor=lightgray,opacity=0.4](1.5,1)(0.13,0.13)
\rput(1,2){\footnotesize $d$} 
\rput(0,1.5){\footnotesize $a$}
\rput(2,2){\footnotesize $e$}
\rput(1.5,1){\footnotesize $b$}
\end{pspicture}
\end{center}
\end{figure}
%
%%%%%%%%%%%%%%%%%%%%%%%%%%%%%%%%%	 
	We check that any extension of $\rhd$ list-rationalizes $c\,$.\vs\vs
	\begin{itemize}
		\item $c(abd) = a\,$: If $\min(abd,\rhd) = a$, then $c(abd) = c(c(bd)a)= c(ab) = a$. If $\min(abd,\rhd) = b$, then $c(abd) = c(c(ad)b)= c(ab) = a$.  
		\item $c(abe) = a\,$: If $\min(abe,\rhd) = a$, then $c(abe) = c(c(be)a)= c(ae) = a$. If $\min(abe,\rhd) = b$, then $c(abe) = c(c(ae)b)= c(ab) = a$.  
		\item $c(ade) = a\,$: If $\min(abd,\rhd) = a$, then $c(ade) = c(c(de)a)= c(ad) = a$. If $\min(ade,\rhd) = d$, then $c(ade) = c(c(ae)d)= c(ad) = a$. If $\min(ade,\rhd) = e$, then $c(ade) = c(c(ad)e)= c(ae) = a$. 
		\item $c(bde) = c(c(de)b) = c(bd) = b$. 
		\item $c(abde) = a\,$: If $\min(abde,\rhd) = a$, then $c(abde) = c(c(bde)a)= c(ab) = a$. If $\min(abde,\rhd) = b$, then $c(abde) = c(c(ade)b)= c(ab) = a$. 
	\end{itemize}

	\item[4:]$c(ab)=a$, $c(ad)=d$, $c(ae)=a$, $c(bd)=b$, $c(be)=b$, $c(de)=d$, $c(abd)=a$, $c(abe)=a$, $c(ade)=d$, $c(bde)=b$, $c(abde)=a$.
	Assume $c$ is LR.
	From $c(ad)=d$ and $c(abd)=a$, we obtain $d\rhd a$ and $b\rhd a$.
	Thus, $\rhd$ must extend the following partial order: 
		%%%%%%%%%%%%  LR Class 3  %%%%%%%%%%%%%%%%
%
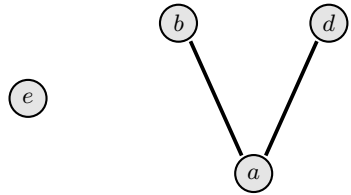
\begin{figure}[H]
\begin{center}
\psset{xunit=2} \psset{yunit=2}  
\begin{pspicture}[showgrid=false](-3,1.3)(5,2.1)  
\psline[arrowsize=6pt,linewidth=0.05](1.08,1.88)(1.42,1.12)
\psline[arrowsize=6pt,linewidth=0.05](1.92,1.88)(1.58,1.12)
\psellipse[fillstyle=solid,fillcolor=lightgray,opacity=0.4](1,2)(0.13,0.13)
\psellipse[fillstyle=solid,fillcolor=lightgray,opacity=0.4](0,1.5)(0.13,0.13)   
\psellipse[fillstyle=solid,fillcolor=lightgray,opacity=0.4](2,2)(0.13,0.13)
\psellipse[fillstyle=solid,fillcolor=lightgray,opacity=0.4](1.5,1)(0.13,0.13)
\rput(1,2){\footnotesize $b$} 
\rput(0,1.5){\footnotesize $e$}
\rput(2,2){\footnotesize $d$}
\rput(1.5,1){\footnotesize $a$}
\end{pspicture}
\end{center}
\end{figure}
%
%%%%%%%%%%%%%%%%%%%%%%%%%%%%%%%%%	 
	We check that any extension of $\rhd$ list-rationalizes $c$.\vs\vs
	\begin{itemize}
		\item $c(abd) = c(c(bd)a) = c(ab) = a$.  
		\item $c(abe) = a\,$: If $\min(abe,\rhd) = a$, then $c(abe) = c(c(be)a)= c(ab) = a$. If $\min(abe,\rhd) = e$, then $c(abe) = c(c(ab)e)= c(ae) = a$.  
		\item $c(ade) = d\,$: If $\min(abd,\rhd) = a$, then $c(ade) = c(c(de)a)= c(ad) = d$. If $\min(ade,\rhd) = e$, then $c(ade) = c(c(ad)e)= c(de) = d$.  
		\item $c(bde) = b\,$: If $\min(bde,\rhd) = b$, then $c(bde) = c(c(de)b)= c(bd) = b$. If $\min(bde,\rhd) = d$, then $c(bde) = c(c(be)d)= c(bd) = b$.   
		\item $c(abde) = a\,$: If $\min(abde,\rhd) = a$, then $c(abde) = c(c(bde)a)= c(ab) = a$. If $\min(abde,\rhd) = e$, then $c(abde) = c(c(abd)e)= c(ae) = a$.\vs 
	\end{itemize}
\end{description}

\noindent Summing up Classes 1-4, out of $11$ RGT choices there are exactly $7 + 1 +1 +1 =10$ LR choices (the only choice that is RGT but not LR is the one in subcase 1E). 
\end{proof}
 
%%%%%%%%%%%%%%%%%%%%%%%%%%%%%%%%%%%
%%%%%%%%%%%%%%%%%%%%%%%%%%%%%%%%%%%

\subsection{Sequentially rationalizable (SR)} %\textbf{(in fieri)}} 

{\large \begin{lemma} \label{LEM:SR}
There are exactly $15$ non-isomorphic SR choices on $X$.
\end{lemma}}

\begin{proof}
Suppose $c$ is SR.
By definition, there is an ordered list $\L=(\succ^{1}, \ldots,\succ^{n})$ of asymmetric relations on $X$ such that the equality $c(A)=M_{n}(A)$ holds for all $ A\in\X$ (where $M_n(A)$ has been defined in Section~\ref{SECT:background}). 

To start, we introduce some compact notation.
For any $x_i,x_i,x_p,x_q\in X$, we write:\vs\vs
\begin{itemize}
	\item $x_i\rightarrowtail  x_j$ (which stands for ``$x_i$ \textsl{eliminates} $x_j$'') if there exists $\succ^{s}\,\in\L$ with the property that $x_i\succ^{s}x_j$, and $\neg(x_i \succ^r x_j \,\vee\, x_j\succ^{r}x_i)$ for any $\succ^{r}\,\in\L$ such that $r<s$;\vs\vs
	\item $(x_i\rightarrowtail  x_j)\,\mathbf{B}\,(x_p\rightarrowtail  x_q)$ (which stands for ``$x_i$ \textsl{eliminates} $x_j$ \textsl{Before} $x_p$ \textsl{eliminates} $x_q$'') if there exist $\succ^{s},\succ^{u}\,\in\L$ with the property that\vs\vs
	\begin{itemize}
		\item $x_i\succ^{s}x_j$ and $\neg(x_i \succ^r x_j \, \vee \, x_j\succ^{r}x_i)$ for any $\succ^{r}\,\in\L$ such that $r<s$,\vs 
		\item $x_p\succ^{u}x_q$ and $\neg(x_p \succ^t x_q \, \vee \, x_q\succ^{t}x_p)$ for any $\succ^{t}\,\in\L$ such that $t<u$, and\vs 
		\item $s<u$.\vs\vs
	\end{itemize}
\end{itemize}
In other words, $x_i \r x_j$ means that there is a rationale $\succ_s$ (with minimum index $s$) in the list $\L = (\succ_1,\succ_2,\ldots,\succ_n)$ which witnesses a strict preference of $x_i$ over $x_j$, and $x_j$ is never preferred to $x_i$ for all rationales $\succ_1,\ldots, \succ_s$.
This implies that if $\L$ sequentially rationalizes $c$, then in a pairwise comparison (but not necessarily in larger menus) $x_i$ is chosen over $x_j$. 

Similarly, $(x_i\rightarrowtail  x_j)\,\mathbf{B}\,(x_p\rightarrowtail  x_q)$ means that if $\L$ sequentially rationalizes $c$, then (in pairwise comparisons) $x_i$ eliminates $x_j$, $x_p$ eliminates $x_q$, and the former process of elimination strictly precedes the latter. 
Note that some of the items $x_i,x_j,x_p,x_q$ maybe be the same (in fact, $x_j = x_p$ will often happen in applications). 
%
%{\blue I'm not sure $\mathbf{B}$ is asymmetric.
%Consider three relations: $\succ^{1},\succ^{2},\succ^{3}$ such that
%\begin{itemize}
%	\item $x_1 \succ^{1} x_2$;
%	\item $x_3 \succ^{2} x_4$;
%	\item $x_1 \succ^{3} x_2$.
%\end{itemize}
%We obtain $(x_1 \r x_2)B(x_3 \r x_4)$ by looking at $\succ^{1}$ and $\succ^{2}$, and $(x_3 \r x_4)B(x_1 \r x_2)$ by looking at $\succ^{2}$ and $\succ^{3}$.
%I presume the definition of $\mathbf{B}$ should state the $s$ and $u$ are the first indexes to eliminate the two items. 
%}
%
The following result is useful:\vs
\begin{lemma}
\label{LEM:SR_constraints} 
Let $x_1,x_2,x_3,x_4\in X$ and $A\subseteq X$.
We have:\vs
\begin{itemize} 
	\item[\rm(i)]  $\rightarrowtail$ is asymmetric and complete;\footnote{{ }A binary relation $R$ on $X$ is \textsl{complete} if for all distinct $x,y \in X$, either $x R y$ or $y R x$ (or both).}\vs 
	\item[\rm(ii)] $\mathbf{B}$ is asymmetric and transitive;\footnote{{ }By the transitivity of $\mathbf B$, we use $(x_1\rightarrowtail  x_2)\,\mathbf{B}\,(x_2\rightarrowtail  x_3)\,\mathbf{B}\,(x_3\rightarrowtail  x_4)$ in place of $(x_1\rightarrowtail  x_2)\,\mathbf{B}\,(x_2\rightarrowtail  x_3)$ and $(x_2\rightarrowtail  x_3)\,\mathbf{B}\,(x_3\rightarrowtail  x_4)$.}\vs  
	\item[\rm(iii)]  $x_1 \r x_2 \;\; \Longleftrightarrow \;\; c(x_1x_2)=x_1$;\vs
	\item[\rm(iv)] $x_1 \r x_2 \,\wedge \, x_1 \r x_3 \;\;\Longrightarrow\;\; c(x_1 x_2 x_3) = x_1$;\vs
	\item[\rm(v)] $c(x_1 x_2 x_3) = x_1 \;\;\Longrightarrow\;\; x_1 \r x_2 \,\vee\, x_1 \r x_3$;\vs
	\item[\rm(vi)] $x_1 \r x_2 \,\wedge \, x_1 \r x_3 \, \wedge\, x_1 \r x_4 \;\;\Longrightarrow\;\; c(x_1 x_2 x_3 x_4) = x_1$;\vs
	\item[\rm(vii)] $c(x_1 x_2 x_3 x_4) = x_1 \;\;\Longrightarrow\;\; x_1 \r x_2  \,\vee\, x_1 \r x_3  \,\vee\, x_1 \r x_4$;\vs
	\item[\rm(viii)]  $c(x_1 x_2)=x_1 \,\wedge\, c(x_1 x_2 x_3)=x_2 \;\; \Longrightarrow \;\; (x_3 \r x_1) \B (x_1 \r  x_2)$;\vs% \wedge \, (x_2 \r x_3)$;
	\item[\rm(ix)]  $c(x_1 x_2)\!=\!x_1 \,\wedge\, c(x_1 x_2 x_3 x_4)\!=\!x_2 \,\Longrightarrow\, (x_3 \r x_1)\mathbf{B}(x_1 \r x_2) \vee (x_4 \r x_1)\mathbf{B}(x_1 \r x_2)$;\vs
	\item[\rm(x)]  $c(x_1 x_2)=x_1 \,\wedge\, c(x_1 x_3)=x_1 \,\wedge\, c(x_1 x_2 x_3 x_4)=x_2 \;\;\Longrightarrow\;\; (x_4 \r x_1) \B (x_1 \r x_2)$;\vs
%\item $\left((x_1\rightarrowtail  x_2)\mathbf{B}(x_2\rightarrowtail  x_3)\right)\wedge \left((x_2\rightarrowtail  x_3)\mathbf{B}(x_3\rightarrowtail  x_4)\right) \Rightarrow \neg\left( (x_3\rightarrowtail  x_4)\mathbf{B}(x_4\rightarrowtail  x_1)\right)$.
	\item[\rm(xi)] $(x_1 \r x_2) \B (x_2 \r x_3) \B (x_3 \r x_1) \;\; \Longrightarrow\;\; c(x_1 x_2 x_3)=x_3${\blue ;}
	\item[\rm(xii)] $c(A)\neq x_1 \implies (\exists r \in \{1,\ldots ,n\})\,(\exists a \in A)\; a \succ^{r} \! x_1 \,\wedge\, a,x \in M_{r-1}(A)$. 
\end{itemize}
\end{lemma} 

\begin{proof}
The proofs of parts (i)--(vii) are straightforward, and are left to the reader.\vs\vs
\begin{itemize} 
\item[\rm(viii)] Toward a contradiction, suppose the antecedent of the implication holds, but the consequent fails. %$c(x_1x_2)=x_1$, $c(x_1x_2x_3)=x_2$, and $\neg\left( (x_3 \r x_1) \B (x_1 \r  x_2) \right)$.
Since $c(x_1x_2)=x_1$, we get $x_1 \r x_2$ by part (iii). 
Furthermore, since $c(x_1 x_2 x_3) \neq x_1$, part (iv) implies that $x_1 \r x_3$ does not hold, hence $x_3 \r x_1$ by part (i). 
Now the hypothesis $\neg\left( (x_3 \r x_1) \B (x_1 \r  x_2) \right)$ implies that $x_3$ eliminates $x_1$ either at the same time or after $x_1$ eliminates $x_2$.
%However, if the two eliminations are simultaneous (i.e., the asymmetric relations witnessing them are the same), then we get $c(x_1 x_2 x_3) = x_3$, a contradiction. 
By way of contradiction, suppose $x_3 \r x_1$ and $x_1 \r x_2$ happen at the same time.
By definition, there is $r \in \{1,\ldots , n\}$ such that $x_3 \succ^{r} \!x_1$ and $x_1 \succ^{r}\! x_2$.
The assumption $c(x_1x_2x_3)=x_2$ together with $x_1 \succ^{r} \! x_2$ implies that $x_1$ must be eliminated before $\succ^{r}$ applies to the menu $x_1 x_2 x_3$.
Therefore, we must have $x_2 \succ^{s} \!x_1$ or $x_3 \succ^{s} \!x_1$ for some $s<r$.
However, we have $\lnot(x_2 \succ^{s} \!x_1)$, because $s<r$ and $x_1 \r x_2$ with $x_1 \succ^{r} \!x_2$.
Hence $x_3 \succ^{s} \!x_1$ for some $s<r$.
We conclude that the elimination was not simultaneous.
It follows that $(x_1 \r x_2) \B (x_3 \r x_1)$. 
By a similar argument, one can derive a contradiction also in this case. 

%Since $c(x_1x_2x_3)=x_2$, we must have $(x_2 \rightarrowtail x_1) \mathbf{B}( x_1 \r x_2)$, which contradicts part (i). 

\item[\rm(ix)] Toward a contradiction, suppose the antecedent of the implication holds, but the consequent fails. 
Since $c(x_1x_2)=x_1$, we get $x_1 \r x_2$ by part (iii). 
Furthermore, since $c(x_1 x_2 x_3 x_4) \neq x_1$, we get $x_3 \r x_1$ or $x_4 \r x_1$ (or both) by part (vi). 
The assumption implies that both $x_3 \r x_1$ and $x_4 \r x_1$ never happen before $x_1 \r x_2$.  
In any case, we get $c(x_1 x_2 x_3 x_4) \neq x_2$, a contradiction.  
%$c(x_1x_2)\!=\!x_1$, $c(x_1x_2x_3x_4)=x_2$ and $\neg\left( (x_3 \r x_1)\mathbf{B}(x_1 \r x_2) \vee (x_4 \r x_1)\mathbf{B}(x_1 \r x_2) \right)$.
 %This means that $x_3$ and $x_4$ can eliminate $x_1$ only at the same time or after $x_1$ eliminates $x_2$.
 %Since $c(x_1x_2x_3x_4)=x_2$, we must have $(x_2 \rightarrowtail x_1) \mathbf{B}( x_1 \r x_2)$, which contradicts
%L\ref{LEM:SR_constraints}(i).

\item[\rm(x)] Toward a contradiction, suppose the antecedent of the implication holds, but the consequent fails. 
By part (iii), we get $x_1 \r x_2$ and $x_1 \r x_3$. 
Furthermore, part (vii) yields $\neg(x_1 \r x_4)$, whence $x_4 \r x_1$ by the completeness of $\r$. 
Since $(x_4 \r x_1) \mathbf B (x_1 \r x_2)$ fails whereas both $x_4 \r x_1$ and $x_1 \r x_2$ hold, it must happen that $x_4$ eliminates $x_1$ simultaneously or after $x_1$ eliminates $x_2$.
Since $c(x_1x_2x_3x_4)=x_2$, there must be $x_i\in x_2 x_3$ such that $(x_i \rightarrowtail x_1) \mathbf{B}( x_1 \r x_2)$, in particular $x_i \r x_1$. 
This is impossible by the asymmetry of $\r$. 

\item[\rm(xi)] If the antecedent holds, then $c(x_1 x_2 x_3)$ must be different from both $x_1$ and $x_2$. The claim follows. 
%Assume $c(x_1x_2x_3)=x_1$.
%Note that $x_1$ eliminates $x_2$ that cannot eliminate $x_3$, which eliminates $x_1$.
%Thus there must be $x_i\in x_1x_2$ such that $ (x_i \r x_3) \B (x_3 \r x_1)$.
%Whatever $x_i$ is, L\ref{LEM:SR_constraints}(i) is contradicted.
%Assume $c(x_1x_2x_3)=x_2$.
%Note that $x_1$ eliminates $x_2$.
%Thus there must be $x_i\in x_2x_3$ such that $(x_i \r x_1) \B (x_1 \r x_2)$.
%Whatever $x_i$ is, L\ref{LEM:SR_constraints}(i) is contradicted.
\item[\rm(xii)] If $c(A)\neq x_1$, then we obtain $x_1 \notin M_r(A)$ for some $r \in \{1,\ldots , n\}$.
Take the minimum $s$ such that $x_1 \notin M_s(A)$.
By definition, $x_1$ was eliminated by some elements in $M_{s-1}(A)\subseteq A$, which is our claim.
\end{itemize}
\vspace{-0.95cm}
\end{proof}

To count SR choices, we employ Approach\:\#1.
As in the proof of Lemma~\ref{LEM:RGT}, the implication `SR$\;\Longrightarrow\;$AC'  \citep{ManziniMariotti2007} comes handy   to simplify the counting. 
Since several deduction will be based on Lemma~\ref{LEM:SR_constraints}, to keep notation compact we use `L\ref{LEM:SR_constraints}(iii)' in place of `Lemma~\ref{LEM:SR_constraints}(iii)', `L\ref{LEM:SR_constraints}(v)' in place of `Lemma~\ref{LEM:SR_constraints}(v)', etc.

\begin{description}
\item[\textsc{Class} 1: (4-cycle):]$c(ab)=a$, $c(ad)=a$, $c(ae)=e$, $c(bd)=b$, $c(be)=b$, and $c(de)=d$.  
Assume $c$ is SR.
By AC, we get $c(abd)=a$ and $c(bde)=b$.
We need to determine $c(abe)$, $c(ade)$, and $c(abde)$. 
According to the three possible selections from the menu $abe$, we distinguish three cases: (1A) $c(abe)=a\,$; (1B) $c(abe)=b\,$; (1C) $c(abe)=e\,$.  
 
\begin{description}
\item[1A:]$c(abe)=a$.

\textsc{Claim:} $c(abde) =a$. Toward a contradiction, assume $c(abde)\neq a$.
By L\ref{LEM:SR_constraints}(xii), there are $x \in X$ and $\succ^r \,\in \L$ such that $x \succ^{r} \!a$ and $x,a \in M_{r-1}(abde)$, whence $x \r a$.
Since $c(ab)=c(ad)=a$, we get $a \r b$ and $a \r d$ by L\ref{LEM:SR_constraints}(iii), hence $x =e$ by the asymmetry of $\r$. 
%Note that $r$ is the minimum integer witnessing $e \r a$, otherwise the condition $a \in  M_{r-1}(abde)$ would not be satisfied.  
By  L\ref{LEM:SR_constraints}(viii), $c(ae)=e$ and $c(abe)=a$ yield $(b \r e) \B (e \r a)$ and $\lnot((a \r b)\mathbf{B}(b \r e))$.
In particular, $e$ is eliminated by $b$ using some rationale $\succ^{s}$ such that $s<r$. 
(Note that since $c(bd) =b$, we have $b \r d$ by L\ref{LEM:SR_constraints}(iii), and so $b$ cannot be eliminated by $d$.)  
This is a contradiction, since $e \in M_{r-1}(abde)$, whereas the last result tells us that $e \notin M_{s}(abde) \supseteq M_{r-1}(abde)$.

%Claims 1--3 yield $c(abde)=a$. 
From the Claim, it follows that 1A generates the following $3$ non-isomorphic choices, which are obtained by considering all possible selections from the menu $ade$ (for simplicity, in each menu we underline the selected item):\footnote{{ }Since this proof will also be used to count choices that are either RSM or CLS, we shall emphasize in magenta all SR choices, in order to facilitate their retrieval by the reader.}\vs
\begin{itemize}  \magenta 
	\item[(1)] \magenta $\u a b,\;\; \u ad,\;\; a \u e,\;\; \u bd,\;\; \u be,\;\; \u de,\;\; \u abd,\;\; \u abe,\;\; \u ade,\;\; \u bde,\;\; \u abde\,$;
	\item[(2)] $\u a b,\;\; \u ad,\;\; a \u e,\;\; \u bd,\;\; \u be,\;\; \u de,\;\; \u abd,\;\; \u abe,\;\; a \u de,\;\; \u bde,\;\; \u abde\,$; 
	\item[(3)] $\u a b,\;\; \u ad,\;\; a \u e,\;\; \u bd,\;\; \u be,\;\; \u de,\;\; \u abd,\;\; \u abe,\;\; a d \u e,\;\; \u bde,\;\; \u abde\,$.
\end{itemize}
To complete our analysis, we check that these choices are sequentially rationalized by a list $\L$ of acyclic (not necessarily transitive) relations: 
\begin{itemize}
	\item[(1)] $\left(\succ^1,\succ^2\right)$, with $a \succ^1 \!b \succ^1 \!d \succ^1 \!e$, $a \succ^1 \!d$, $b \succ^1 \!e$, and $e \succ^2 \!a\,$;
	%$b,d \succ^1 e$, $e\succ^2a$, and $a\succ^3 b \succ^3 d \succ^3 e$ ($\succ^3$ linear order);
	\item[(2)] $\left(\succ^1,\succ^2,\succ^3\right)$, with $b \succ^1 \!e$, $e \succ^2 \!a$, $a \succ^3 \!b \succ^3 \!d \succ^3 \!e$, and $\succ^3$ transitive;\footnote{{ }Note that no list with two rationales suffices. Indeed, this choice is not RSM, because WWARP fails, since $c(ad) =a = c(abde)$ and yet $c(ade)=d$.}
	\item[(3)] $\left(\succ^1,\succ^2\right)$, %$a\succ^1 d$,$b\succ^2 d$,$b\succ^1 e$, $a\succ^2 b , d \succ^2 e$, $e\succ^2 a$, where $\succ^3$ is transitive.
	with $a \succ^1 \!d$, $b\succ^1 \!d$, $b\succ^1 \!e$, and $d \succ^2 \! e \succ^2 \!a \succ^2 \!b\,$. 
\end{itemize}
\medskip

\item[1B:]$c(abe)=b$.
Since $c(ab)=a$, we get $(e \r a) \B (a \r b)$ by L\ref{LEM:SR_constraints}(viii).
We distinguish $3$ subcases (i), (ii), and (iii), according to the choice on $ade$. 

\begin{description}
\item[(i):]$c(ade)=a$.
Since $c(ae)=e$, we have $(d \r e) \B (e \r a)$ by L\ref{LEM:SR_constraints}(viii).
Thus, we obtain the chain $(d \r e) \B (e \r a) \B (a \r b)$.
It is not difficult to show that $c(abde) \neq d,e$. 
It follows that only two choices need be checked, namely
\begin{itemize} \magenta
	\item[(4)] $\u a b,\;\; \u ad,\;\; a \u e,\;\; \u bd,\;\; \u be,\;\; \u de,\;\; \u abd,\;\; a \u be,\;\; \u ade,\;\; \u bde,\;\; \u abde\,$;
	\item[(5)] $\u a b,\;\; \u ad,\;\; a \u e,\;\; \u bd,\;\; \u be,\;\; \u de,\;\; \u abd,\;\; a \u be,\;\; \u a de,\;\; \u bde,\;\; a \u b de\,$.
\end{itemize}
Both choices are sequentially rationalized by a list $\L$ as follows:  
\begin{itemize}
	\item[(4)] $\left(\succ^1,\succ^2,\succ^3 \right)$, with $d\succ^1 \!e$, $e\succ^2 \!a$, $a\succ^3 \!b \succ^3 \!d \succ^3 \!e$, and $\succ^3$ transitive;\footnote{{ }Since $(d\rightarrowtail  e)\mathbf{B}(e\rightarrowtail  a)\mathbf{B}(a\rightarrowtail  b)$ holds, $c$ is not RSM. In fact, WWARP fails.}
	\item[(5)] $\left(\succ^1,\succ^2,\succ^3,\succ^4\right)$, with $b \succ^1 \!d$, $d \succ^2 \!e$, $e \succ^3\!a$, $a \succ^4 \! b \succ^4 \!e$, and $a \succ^4\! d$.\footnote{{ }Since $(b\rightarrowtail  d)\mathbf{B}(d\rightarrowtail  e)\mathbf{B}(e\rightarrowtail  a)\mathbf{B}(a\rightarrowtail  b)$ holds, $c$ is not RSM (and not even SR by $3$ rationales).}  
\end{itemize}
%
%For choice 5, the sequence for $\B$ is obtained as follows. 
%Since $(d\rightarrowtail  e)\mathbf{B}(e\rightarrowtail  a)\mathbf{B}(a\rightarrowtail  b)$, 
%and there is no item other that $e$ that can eliminate $a$, there must be $x_i\in abe$  such that $(x_i\rightarrowtail  d)\mathbf{B}(d\rightarrowtail  e)\mathbf{B}(e\rightarrowtail  a)\mathbf{B}(a\rightarrowtail  b)$.
%If $x_i=a$, L\ref{LEM:SR_constraints}(viii) yields $c(ade)=e$, which is false.
%If $x_i=e$, L\ref{LEM:SR_constraints}(i) is violated.
%Thus $x_i=b$, and we obtain $(b\rightarrowtail  d)\mathbf{B}(d\rightarrowtail  e)\mathbf{B}(e\rightarrowtail  a)\mathbf{B}(a\rightarrowtail  b)$.
%

\item[(ii):]$c(ade)=d$.
Since $c(ad)=a$, we get $(e\rightarrowtail  a)\mathbf{B}(a\rightarrowtail  d)$ by L\ref{LEM:SR_constraints}(viii). 
We already know that $(e\rightarrowtail  a)\mathbf{B}(a\rightarrowtail  b)$.
An argument similar to that used in the previous cases yields $c(abde) =b$.
Thus, the only feasible choice $c$ is
\begin{itemize} \magenta 
	\item[(6)] $\u a b,\;\; \u ad,\;\; a \u e,\;\; \u bd,\;\; \u be,\;\; \u de,\;\; \u abd,\;\;  a\u be,\;\;  a\u de,\;\; \u bde,\;\;  a\u bde\,$.
\end{itemize}
This choice is SR, and a rationalizing list $\L$ is the following:
\begin{itemize}
	\item[(6)] $\left(\succ^1,\succ^2 \right)$, with $e \succ^1 \!a$, $a\succ^2 \! b \succ^2 \! d \succ^2 \! e$, and $\succ^2$ transitive.
\end{itemize}

\item[(iii):]$c(ade)=e$.
Since $c(de)=d$, we get $(a\rightarrowtail  d)\mathbf{B}(d\rightarrowtail  e)$ by L\ref{LEM:SR_constraints}(viii).
We already know that $(e\rightarrowtail  a)\mathbf{B}(a\rightarrowtail  b)$. 
As in subcase (ii), we get $c(abde) =b$. 
Thus, $c$ is defined as follows:
\begin{itemize} \magenta 
	\item[(7)] $\u a b,\;\; \u ad,\;\; a \u e,\;\; \u bd,\;\; \u be,\;\; \u de,\;\; \u abd,\;\;  a\u be,\;\;  ad\u e,\;\; \u bde,\;\;  a\u bde\,$.
\end{itemize}
This choice is SR, and a rationalizing list $\L$ is the following: 
\begin{itemize}
	\item[(7)] $\left(\succ^1,\succ^2\right)$, with $e \succ^1 \! a \succ^1 \!d$, $a\succ^2 \! b \succ^2 \! d \succ^2 \! e$, and $\succ^2$ transitive.
\end{itemize}
\end{description}

\item[1C:]$c(abe)=e$.
Since $c(be)=b$, we get $(a \r b) \B (b \r e)$ by L\ref{LEM:SR_constraints}(viii).
We claim that $c(abde)\neq b$.
Otherwise, $c(ab)=a$ and $c(ad)=a$ yield $(e \r a) \B (a \r b)$ by L\ref{LEM:SR_constraints}(x), whence the chain $(e \r a) \B (a \r b) \B (b \r e)$ implies $c(abe) = b$ by L\ref{LEM:SR_constraints}(x), which is false.  
Thus, there are $3$ subcases, according to the choice on $ade$.

\begin{description}
\item[(i):]$c(ade)=a$.
Since $c(ae)=e$, we get $(d \r e) \B (e \r a)$ by L\ref{LEM:SR_constraints}(viii).
It is simple to prove $c(abde)\neq d$, hence $c(abde) \neq b,d$. 
%
%\textsc{Claim~13:} $c(abde)\neq d$.
%Assume toward a contradiction that $c(abde)=d$.
%Since $c(ad)=a$ and $c(ab)=a$, Lemma~\ref{LEM:SR_constraints}(v) would imply that $(e\rightarrowtail  a)\mathbf{B}(a\rightarrowtail  d)$.
%Since $c(bd)=b$ and $c(be)=b$, Lemma~\ref{LEM:SR_constraints}(v) would imply that $(a\rightarrowtail  b)\mathbf{B}(b\rightarrowtail  d)$.
%Thus, we obtain the chain $(d\rightarrowtail  e)\mathbf{B}(e\rightarrowtail  a)\mathbf{B}(a\rightarrowtail  d)$.
%In this chain $d$ eliminates $e$ that cannot eliminate $a$, which eliminates $e$. 
%Since any element other than $a$ cannot eliminate $b$, we conclude that $c(abde)\neq d$.
%
It follows that only two choices need be checked:
\begin{itemize} \magenta 
	\item[(8)] $\u a b,\;\; \u ad,\;\; a \u e,\;\; \u bd,\;\; \u be,\;\; \u de,\;\; \u abd,\;\;  ab\u e,\;\;  \u ade,\;\; \u bde,\;\;  \u abde\,$;
	\item[(9)] $\u a b,\;\; \u ad,\;\; a \u e,\;\; \u bd,\;\; \u be,\;\; \u de,\;\; \u abd,\;\;  ab\u e,\;\;  \u ade,\;\; \u bde,\;\;   abd\u e\,$.
\end{itemize}
Both choices are sequentially rationalized by a list $\L$ as follows:  
\begin{itemize}
	\item[(8)] $\left(\succ^1,\succ^2 \right)$, with $a \succ^1 \!b$, $d \succ^1 \!e$, $b \succ^2 \! e \succ^2 \! a \succ^2 d$, and $\succ^2$ transitive;
	\item[(9)] $(\succ^1,\succ^2,\succ^3)$, with $b\succ^1 \! d$, $a \succ^2 \! b$, $a \succ^2 \! d \succ^2 \!e$, $b \succ^3 \!e$, and $d\succ^3 e\succ^3 a\,$.\footnote{{ }Since $(b \r d) \B (d \r e) \B (e \r a)$ holds, $c$ is not RSM. Note that WWARP fails, because $c(ae)=e=c(abde)$ and yet $c(ade)=a$.} 
\end{itemize}
%
%For choice 9: Since $c(de)=d$, L\ref{LEM:SR_constraints}(iv) implies that $(a\rightarrowtail  d)\mathbf{B}(d\rightarrowtail  e)$ or $(b\rightarrowtail  d)\mathbf{B}(d\rightarrowtail  e)$, or both.
%If $(a\rightarrowtail  d)\mathbf{B}(d\rightarrowtail  e)$ holds, we obtain the chain $(a\rightarrowtail  d)\mathbf{B}(d\rightarrowtail  e)\mathbf{B}(e\rightarrowtail  a)$.
% L\ref{LEM:SR_constraints}(viii) implies $c(ade)=e$, which is false.
%We conclude that only $(b\rightarrowtail  d)\mathbf{B}(d\rightarrowtail  e)$ holds.
%Thus, we have $(b\rightarrowtail  d)\mathbf{B}(d\rightarrowtail  e)\mathbf{B}(e\rightarrowtail  a)$.
%

\item[(ii):]$c(ade)=d$.
Since $c(ad)=a$, we get $(e \r a) \B (a \r d)$ by L\ref{LEM:SR_constraints}(viii).
It is simple to prove $c(abde)\neq a$, hence $c(abde) \neq a,d$. 
%
% \textsc{Claim~15:} $c(abde)\neq a$.
%Toward a contradiction assume that $c(abde)=a$.
%Since $c(ae)=e$, L\ref{LEM:SR_constraints}(iv) implies that
%there is $x_i\in bd$ such that $(x_i\rightarrowtail  e)\mathbf{B}(e\rightarrowtail  a)$.
%If $x_i=d$, we obtain the sequence $(d\rightarrowtail  e)\mathbf{B}(e\rightarrowtail  a)\mathbf{B}(a\rightarrowtail  d)$.
%L\ref{LEM:SR_constraints}(viii) implies $c(ade)=a$, which is false.
%Thus, assume $x_i=b$.
%We get the sequence
% $(b\rightarrowtail  e)\mathbf{B}(e\rightarrowtail  a)\mathbf{B}(a\rightarrowtail  d)$.
%Since we already know that in case 1C $(a\rightarrowtail  b)\mathbf{B}(b\rightarrowtail  e)$ always holds, we conclude that $(a\rightarrowtail  b)\mathbf{B}(b\rightarrowtail  e)\mathbf{B}(e\rightarrowtail  a)\mathbf{B}(a\rightarrowtail  d)$.
%In this sequence $a$ is eliminated, implying $c(abde)\neq a$, which is false.
%
It follows that only two choices need be checked:
\begin{itemize} \magenta 
	\item[(10)] $\u a b,\;\; \u ad,\;\; a \u e,\;\; \u bd,\;\; \u be,\;\; \u de,\;\; \u abd,\;\;  ab\u e,\;\;  a \u de,\;\; \u bde,\;\; ab \u de\,$;
	\item[(11)] $\u a b,\;\; \u ad,\;\; a \u e,\;\; \u bd,\;\; \u be,\;\; \u de,\;\; \u abd,\;\;  ab\u e,\;\;  a\u de,\;\; \u bde,\;\;   abd\u e\,$.
\end{itemize}
Both choices are sequentially rationalized by a list $\L$ with two rationales:  
\begin{itemize}
	\item[(10)] $e \succ^1\! a \succ^1 \! b$, $a \succ^2 \! b \succ^2 \! d \succ^2 \! e$, and $\succ^2$ transitive; 
	\item[(11)]  $e \succ^1 \! a \succ^1 \! b \succ^1 \!d$, $a\succ^2 \!d \succ^2 \!e$, and $b\succ^2 \!e$.
\end{itemize}
%
% Other counting for choice 11, to determine the structure fo the rationales:
%
% Lemma~\ref{LEM:SR_constraints}(iii) implies that $(a\rightarrowtail  b)\mathbf{B}(b\rightarrowtail  e)$.
%Since $c(de)=d$,  L\ref{LEM:SR_constraints}(iv) yields that $(a\rightarrowtail  d)\mathbf{B}(d\rightarrowtail  e)$ or $(b\rightarrowtail  d)\mathbf{B}(d\rightarrowtail  e)$, or both hold.
%If $(a\rightarrowtail  d)\mathbf{B}(d\rightarrowtail  e)$ holds,  we obtain the sequence  $(e\rightarrowtail  a)\mathbf{B}(a\rightarrowtail  d)\mathbf{B}(d\rightarrowtail  e)$, in which $e$ is eliminated.
%%Since $c(abde)=e$, there must be $x_i\in bd$ such that $(x_i\rightarrowtail  e)\mathbf{B}(e\rightarrowtail  a)\mathbf{B}(a\rightarrowtail  d)\mathbf{B}(d\rightarrowtail  e)$.
%Thus we must have $(b\rightarrowtail  d)\mathbf{B}(d\rightarrowtail  e)$ (and the chain $(e\rightarrowtail  a)\mathbf{B}(a\rightarrowtail  d)\mathbf{B}(d\rightarrowtail  e)$ does not necessary hold).
%
 
\item[(iii):]$c(ade)=e$.
Since $c(de)=d$, we get $(a \r d) \B (d \r e)$ by L\ref{LEM:SR_constraints}(viii).
It is simple to prove $c(abde)\neq a,d$, hence $c(abde) =e$. 
%
% \textsc{Claim~16:} $c(abde)\neq a$.
%Assume toward a contradiction that $c(abde)=a$.
%Since $c(ae)=e$, Lemma~\ref{LEM:SR_constraints}(iv) implies that there is $x_i\in bd$ such that  $(x_i\rightarrowtail  e)\mathbf{B}(e\rightarrowtail  a)$.
%If $x_i=b$, we would have that $
%(b\rightarrowtail  e)\mathbf{B}(e\rightarrowtail  a)$.
%We already know that in case 1C $
%(a\rightarrowtail  b)\mathbf{B}(b\rightarrowtail  e)$ always hold.
%Thus we obtain the chain $
%(a\rightarrowtail  b)\mathbf{B}(b\rightarrowtail  e)\mathbf{B}(e\rightarrowtail  a)$.
%In this sequence $a$ is eliminated, contradicting $c(abde)=a$.
%Thus, we must have $x_i=d$. 
%We get the sequence $(a\rightarrowtail  d)\mathbf{B}(d\rightarrowtail  e)\mathbf{B}(e\rightarrowtail  a)$.
%Note that $a$ is eliminated in this sequence by $e$.
%Item $e$ is not eliminated by $d$ because $a$ eliminates $d$.
%Moreover, we have already shown that $e$ is not eliminated by $b$.
%We conclude $c(abde)\neq a$, which is false. 
%
% \textsc{Claim~17:}  $c(abde)\neq d$.
%Assume toward a contradiction that $c(abde)=d$.
%Since $c(ad)=a$ and $c(ab)=a$, Lemma~\ref{LEM:SR_constraints}(v) would imply that $(e\rightarrowtail  a)\mathbf{B}(a\rightarrowtail  d)$.
%We obtain the sequence $(e\rightarrowtail  a)\mathbf{B}(a\rightarrowtail  d)\mathbf{B}(d\rightarrowtail e)$.
%Lemma~\ref{LEM:SR_constraints}(v) implies that $c(ade)=d$, which is false.
%

Thus, the only feasible choice $c$ is
\begin{itemize}  \magenta 
	\item[(12)] $\u a b,\;\; \u ad,\;\; a \u e,\;\; \u bd,\;\; \u be,\;\; \u de,\;\; \u abd,\;\;  ab \u e,\;\;  ad \u e,\;\; \u bde,\;\;  abd \u e\,$.
\end{itemize}
This choice is SR by a list $\L$ with two rationales: 
\begin{itemize}
	\item[(12)] $\left(\succ^1,\succ^2 \right)$, with $a\succ^1 \!b$, $a \succ^1\! d$, $b\succ^2 \!d \succ^2\! e \succ^2 \! a$, and $\succ_2$ transitive.\vs\vs 
\end{itemize}
 \end{description}
\end{description}

Summarizing, in Class~1 there are $12$ non-isomorphic SR choices.

%%%%%%%%%%%%%%%%%%%%%%%%%%%%%%%%%%%%%%%

\item[\textsc{Class 2} (source and sink):]$c(ab)\!=\!a$, $c(ad)\!=\!a$, $c(ae)\!=\!a$, $c(bd)\!=\!b$, $c(be)\!=\!b$, $c(de)\!=\!d$.
%As usual, we assume $c$ is SR, determine some necessary conditions, and finally check that these conditions are also sufficient to be SR.
Suppose $c$ is SR. 
By AC, we get $c(abd)=c(abe)=c(ade)=c(abde)=a$, and $c(bde)=b$.
Thus, the unique possible SR choice is this class is given by\vs\vs
\begin{itemize} \magenta 
	\item[(13)] $\u a b,\;\; \u ad,\;\; \u a e,\;\; \u bd,\;\; \u be,\;\; \u de,\;\; \u abd,\;\;  \u abe,\;\;  \u ade,\;\; \u bde,\;\; \u abde\,$.\vs\vs
\end{itemize}
This choice is rationalizable, and so it is SR.
 
%%%%%%%%%%%%%%%%%%%%%%%%%%%%%%%%%%%%%% 
 
\item[\textsc{Class} 3 (source but no sink):] $c(ab)=a$, $c(ad)=a$, $c(ae)=a$, $c(bd)=b$, $c(be)=e$, $c(de)=d$.
 Assume $c$ is SR.
 By AC, we get $c(abd)=c(abe)=c(ade)=c(abde)=a$.
 The only remaining menu is $bde$, for which we can assume loss of generality that $c(bde)=b$ (because the other two possibilities $c(bde)=d$ and $c(bde)=e$ yield isomorphic choices). 
 Thus, $c$ is defined by\vs\vs
\begin{itemize} \magenta 
	\item[(14)] $\u a b,\;\; \u ad,\;\; \u a e,\;\; \u bd,\;\; b \u e,\;\; \u de,\;\; \u abd,\;\;  \u abe,\;\;  \u ade,\;\; \u bde,\;\; \u abde\,$.\vs\vs
\end{itemize}
This choice is SR by a list $\L$ with two rationales:\vs\vs 
\begin{itemize}
	\item[(14)] $\left(\succ^1,\succ^2\right)$, with $d \succ^1 \!e$, $a \succ^2 \! e \succ^2 \! b \succ^2 \! d$, and $\succ^2$ transitive. 
\end{itemize}

%%%%%%%%%%%%%%%%%%%%%%%%%%%%%%%%%%%%%%

\item[\textsc{Class} 4 (sink but no source):] $c(ab)=a$, $c(ad)=d$, $c(ae)=a$, $c(bd)=b$, $c(be)=b$, $c(de)=d$.
If $c$ is SR, then $c(abe)=a$, $c(ade)=d$, and $c(bde)=b$ by AC.
Without loss of generality, we can assume $c(abd)=a$ (because the other two possibilities yield isomorphic choices). 
By an argument similar to those described in the previous cases, one can show that $c(abde)=a$.
%	
%	Since $c(ad)=d$ and $c(abd)=a$, Lemma~\ref{LEM:SR_constraints}(iii) implies that  $(b\rightarrowtail  d)\mathbf{B}(d\rightarrowtail  a)$.
%	
%	\textsc{Claim~18:} $c(abde)\neq b$.
%	Assume toward a contradiction that $c(abde)= b$. 
%	Since $c(ab)=a$ and $c(ae)=a$, Lemma~\ref{LEM:SR_constraints}(v) implies that $(d\rightarrowtail  a)\mathbf{B}(a\rightarrowtail  b)$.
%	Thus, we obtain the chain $(b\rightarrowtail  d)\mathbf{B}(d\rightarrowtail  a)\mathbf{B}(a\rightarrowtail  b)$.
%	Note that $b$ is eliminated by $a$, which is not eliminated by $d$, because $d$ is eliminated by $b$.
%	Item $a$ cannot be eliminated by any item other than $d$.
%	We conclude that $c(abde)\neq b$, which is false.
%	
%	\textsc{Claim~19:} $c(abde)\neq d$.
%	Assume toward a contradiction that $c(abde)=d$.
%	Since $c(bd)=b$, and $c(be)=b$ Lemma~\ref{LEM:SR_constraints}(v) implies that $(a\rightarrowtail  b)\mathbf{B}(b\rightarrowtail  d)$.
%	We obtain the sequence $(a\rightarrowtail  b)\mathbf{B}(b\rightarrowtail  d)\mathbf{B}(d\rightarrowtail  a)$.
%	L\ref{LEM:SR_constraints}(v) implies that $c(abd)=d$, which is false.
%	%If $x_i=e$, since $c(be)=b$, we would have that $(e\rightarrowtail  b)$ and $ (b\rightarrowtail  e)$, which is impossible by Lemm.
%	
%	
%	\textsc{Claim~20:}$c(abde)\neq e$.
%	Apply Lemma \ref{LEM:SR_constraints}(vii).
%
%Since $c(ad)=d$ and $c(de)=d$,  implies that $(b\rightarrowtail  d)\mathbf{B}(d\rightarrowtail  a)$.
Thus, there is a unique possible SR choice in this class, and its definition is\vs\vs 
\begin{itemize} \magenta 
	\item[(15)] $\u a b,\;\;  a\u d,\;\; \u ae,\;\; \u bd,\;\; \u be,\;\; \u de,\;\; \u abd,\;\;  \u abe,\;\;   a\u de,\;\; \u bde,\;\;   \u abde\,$.\vs\vs
\end{itemize}
This choice is SR by a list $\L$ with two rationales:\vs\vs
\begin{itemize}
	\item[(15)] $\left( \succ^1,\succ^2 \right)$, with $b\succ^1 \! d$, $d \succ^2 \! a \succ^2 \! b \succ^2 \! e$, and $\succ^2$ transitive. 
\end{itemize}
\end{description} 

\noindent We conclude that there are $15$ non-isomorphic SR choices on $X$, as claimed. 
\end{proof}

%%%%%%%%%%%%%%%%%%%%%%%%%%%%%%%%%%%%%%
%%%%%%%%%%%%%%%%%%%%%%%%%%%%%%%%%%%%%%

\subsection{Status quo bias (SQB)}

{\large \begin{lemma} \label{LEM:SQB}
	There are exactly $6$ non-isomorphic SQB choices on $X$.
\end{lemma}	}
 
\begin{proof}
\cite{ApesteguiaBallester2013} prove that SQB implies SR.
Thus, it suffices to determine which of the 15 SR choices described in Lemma~\ref{LEM:SR} satisfy SQB.
We use the same numeration of cases as in Lemma~\ref{LEM:SR}.\vs\vs
%We introduce some notation, derived from \cite{ApesteguiaBallester2013}.
%We denote by $\mathcal{B}\subset \X$ the family of all binary subsets of $X$.
%We denote by $>$ a linear order (\textsl{route}) over $\mathcal{B}$.
%For any  $A\subseteq X$, we denote by $\mathscr{A}$ the family of any non-empty subset of $A$. 
%Given $A\subseteq X$, we denote by $A^>:=\{B\in\mathcal{B}\cap \mathscr{A} \,\vert\,  B>B^{\prime}\;\;\forall\,B^{\prime}\in\mathcal{B}\cap \mathscr{A} \}$.
%For any $x_1,x_2,x_3,x_4\in X$, we denote by $x_1x_2\cup x_2x_3$ a \textsl{binary partition} of $X$.
%We say that $>$ \textsl{respects} $x_1x_2\cup x_2x_3$ if $x_1x_2< x_1x_3$, $x_1x_2<x_1x_4$  
%
\begin{itemize}
	\item[(1)] $\u a b,\;\; \u ad,\;\; a \u e,\;\; \u bd,\;\; \u be,\;\; \u de,\;\; \u abd,\;\; \u abe,\;\; \u ade,\;\; \u bde,\;\; \u abde$.\vs 
	
	This choice is WSQB: set $a\rhd b\rhd d\rhd e$, $z:=e$, and $Q:=bd$.\vs\vs

\item[(2)] $\u a b,\;\; \u ad,\;\; a \u e,\;\; \u bd,\;\; \u be,\;\; \u de,\;\; \u abd,\;\; \u abe,\;\; a \u de,\;\; \u bde,\;\; \u abde$.\vs

The reader can check that this choice is not SQB.\vs\vs

\item[(3)] $\u a b,\;\; \u ad,\;\; a \u e,\;\; \u bd,\;\; \u be,\;\; \u de,\;\; \u abd,\;\; \u abe,\;\; a d \u e,\;\; \u bde,\;\; \u abde$.\vs 

The reader can check that this choice is not SQB.\vs\vs

\item[(4)] $\u a b,\;\; \u ad,\;\; a \u e,\;\; \u bd,\;\; \u be,\;\; \u de,\;\; \u abd,\;\;  a\u be,\;\; \u ade,\;\; \u bde,\;\; \u abde$.\vs

The reader can check that this choice is not SQB.\vs\vs

\item[(5)] $\u a b,\;\; \u ad,\;\; a \u e,\;\; \u bd,\;\; \u be,\;\; \u de,\;\; \u abd,\;\;  a\u be,\;\; \u ade,\;\; \u bde,\;\;  a\u bde$.\vs

The reader can check that this choice is not SQB.\vs\vs

\item[(6)]$\u a b,\;\; \u ad,\;\; a \u e,\;\; \u bd,\;\; \u be,\;\; \u de,\;\; \u abd,\;\;  a\u be,\;\;  a\u de,\;\; \u bde,\;\;  a\u bde$.\vs

This choice is both ESQB and WSQB: for ESQB, set $a\rhd b\rhd d \rhd e$, $z:=e$, and $Q:=bd$; for WSQB, set $b\rhd d \rhd e\rhd a$, $z:=a$, and $Q:=e$.\vs\vs

\item[(7)] $\u a b,\;\; \u ad,\;\; a \u e,\;\; \u bd,\;\; \u be,\;\; \u de,\;\; \u abd,\;\;  a\u be,\;\;  ad\u e,\;\; \u bde,\;\;  a\u bde$.\vs

The reader can check that this choice is not SQB.\vs\vs

\item[(8)] $\u a b,\;\; \u ad,\;\; a \u e,\;\; \u bd,\;\; \u be,\;\; \u de,\;\; \u abd,\;\;  ab\u e,\;\;  \u ade,\;\; \u bde,\;\;  \u abde$.\vs

The reader can check that this choice is not SQB.\vs\vs

\item[(9)] $\u a b,\;\; \u ad,\;\; a \u e,\;\; \u bd,\;\; \u be,\;\; \u de,\;\; \u abd,\;\;  ab\u e,\;\;  \u ade,\;\; \u bde,\;\;   abd\u e$.\vs

The reader can check that this choice is not SQB.\vs\vs

\item[(10)] $\u a b,\;\; \u ad,\;\; a \u e,\;\; \u bd,\;\; \u be,\;\; \u de,\;\; \u abd,\;\;  ab\u e,\;\;   a\u de,\;\; \u bde,\;\;   ab\u de$.\vs

The reader can check that this choice is not SQB.\vs\vs

\item[(11)] $\u a b,\;\; \u ad,\;\; a \u e,\;\; \u bd,\;\; \u be,\;\; \u de,\;\; \u abd,\;\;  ab\u e,\;\;   a\u de,\;\; \u bde,\;\;   abd\u e$.\vs

The reader can check that this choice is not SQB.\vs\vs

\item[(12)] $\u a b,\;\; \u ad,\;\; a \u e,\;\; \u bd,\;\; \u be,\;\; \u de,\;\; \u abd,\;\;  ab\u e,\;\;   ad\u e,\;\; \u bde,\;\;   abd\u e.$\vs

This choice is ESQB: set $b\rhd d \rhd e\rhd a$, $z:=a$, and $Q:=e$.\vs\vs

\item[(13)] $\u a b,\;\; \u ad,\;\; \u ae,\;\; \u bd,\;\; \u be,\;\; \u de,\;\; \u abd,\;\;  \u abe,\;\;   \u ade,\;\; \u bde,\;\;   \u abde.$\vs

This choice is rationalizable, hence it is SQB.\vs\vs

\item[(14)] $\u a b,\;\; \u ad,\;\; \u ae,\;\; \u bd,\;\;  b\u e,\;\; \u de,\;\; \u abd,\;\;  \u abe,\;\;   \u ade,\;\; \u bde,\;\;  \u abde.$\vs

This choice is both ESQB and WSQB: for ESQB, set $a\rhd e \rhd b \rhd d$, $z:=d$, and $Q:=ab$; for WSBQ, set $a\rhd b\rhd d\rhd e$, $z:=e$, and $Q:=ad$.\vs\vs

\item[(15)] $\u a b,\;\;  a\u d,\;\; \u ae,\;\; \u bd,\;\; \u be,\;\; \u de,\;\; \u abd,\;\;  \u abe,\;\;   a\u de,\;\; \u bde,\;\; \u abde.$\vs

This choice is ESQB: set $d\rhd a\rhd e\rhd b$, $z:=b$, and $Q:=a$.\vs	
\end{itemize}
 Summing up Classes 1--4, there are $3+1+1+1=6$ non-isomorphic SQB choices.
 \end{proof}

%%%%%%%%%%%%%%%%%%%%%%%%%%%%% 
\subsection{Rational shortlist method (RSM)}

{\large \begin{lemma} \label{LEM:RSM}
	There are exactly $11$ non-isomorphic RSM choices on $X$.
\end{lemma}	}
 
 \begin{proof}
 	The claim readily follows from the observations that RSM implies SR, and only 4 of 15 SR choices --namely those numbered (2), (4), (5), and (9), using the numeration in the proof of Lemma~\ref{LEM:SR}-- cannot be rationalized by two asymmetric binary relations.
 \end{proof}
%%%%%%%%%%%%%%%%%%%%%%%%%%%%%%%%%%%
%%%%%%%%%%%%%%%%%%%%%%%%%%%%%%%%%%%

\subsection{Choice by lexicographic semiorders (CLS)}

{\large \begin{lemma} \label{LEM:CLS}
	There are exactly $15$ non-isomorphic CLS choices on $X$.
\end{lemma}	}
 
 \begin{proof}
The claim readily follows from the observation that CLS implies SR, and all 15 SR choices exhibited in the proof of Lemma \ref{LEM:SR} are rationalized by acyclic relations. 
 \end{proof}

Note that the equality between the number of SR and RSM choices on $4$ item is only due to the size of $X$, because on larger ground sets there are choices that are SR but not CLS \citep[Appendix]{ManziniMariotti2012a}.

%%%%%%%%%%%%%%%%%%%%%%%%%%%%%%%%%%%
%%%%%%%%%%%%%%%%%%%%%%%%%%%%%%%%%%%

\subsection{Weak WARP (WWARP)}

{\large \begin{lemma} \label{LEM:WWARP}
There are exactly $304$ non-isomorphic WWARP choices on $X$.
\end{lemma}}

\begin{proof}
%Let the points be $a,b,c,d$ where
%$(a)bcd$, $(b)cd$ and $(c)d$. There are 864 of these choice functions.
We employ Approach\,\#2 to count all choices on $X$ that do \textit{not} satisfy WWARP. 
Suppose $c(abde)=a$, $c(bde)=b$, and $c(de)=d$.
%WARP fails if and only if there are $a,b \in A \subset B\in \X$ such that $c(a,b) = a$, $c(B) = a$, but $c(A) = b$.
WWARP fails if and only if there are two distinct items $x,y \in X$ and two menus $A,B \subseteq X$ such that $x,y \in A \subseteq B$, $c(xy)= c(B)=x$, and yet $c(A) =y$.  
Since $c(X) = c(abde)=a$, WWARP fails if and only if there are $y \in bde$ and $A \subseteq X$ of size $3$ such that $c(ay) =a \in A$ but $c(A) = y$. 
We enumerate all possible cases for the item $y \in bde$, and the menu $A \subseteq X$ containing $a$ and $y$.\vs\vs 
\begin{itemize}
	\item[(1)] $y$ is $b$, and $A$ is either $abd$ or $abe$. Thus, there are two subcases:\vs\vs\vs 
	\begin{itemize}
		\item[(1.i)] $c(ab)=a$ and $c(abd)= b$;\vs\vs 
		\item[(1.ii)] $c(ab) =a$ and $c(abe)=b$.\vs\vs
	\end{itemize}
	\item[(2)] $y$ is $d$, and $A$ is either $abd$ or $ade$. Thus, there are two subcases:\vs\vs\vs 
	\begin{itemize}
		\item[(2.i)] $c(ad)=a$ and $c(abd)= d$;\vs\vs 
		\item[(2.ii)] $c(ad) =a$ and $c(ade)=d$.\vs\vs
	\end{itemize}
	\item[(3)] $y$ is $e$, and $A$ is either $abe$ or $ade$. Thus, there are two subcases:\vs\vs\vs 
	\begin{itemize}
		\item[(3.i)] $c(ae)=a$ and $c(abe)= e$;\vs\vs 
		\item[(3.ii)] $c(ae) =a$ and $c(ade)=e$.\vs\vs
	\end{itemize}
\end{itemize}
Note that these cases may overlap.

Consider now the choice on the menu $ab$, $ad$, and $ae$. 
There are exactly four mutually exclusive cases (I)--(IV). 
In each of them, we count non-WWARP choices.\vs\vs

\begin{description} 
\item[(I)]
Exactly one of $c(ab)=a$, $c(ad)=a$, and $c(ae)=a$ holds. 
This happens for a total of $\frac{3}{8}\, 864=324$ non-isomorphic choices on $X$.
Without loss of generality, assume only $c(ab)=a$ holds (which happens for $\frac{1}{8}\, 864=108$ non-isomorphic choices on $X$).
Now WWARP fails if and only if (1.i) or (1.ii) or both hold, which is true for $\frac{5}{9}\,108=60$ choices.
The same happens when only $c(ad)=a$ holds, or only $c(ae)=a$ holds.
Thus, we get a total of $180$ non-WWARP choices.

\item[(II)] 
Exactly two of $c(ab)=a$, $c(ad)=a$, and $c(ae)=a$ hold.
This happens for a total of $\frac{3}{8}\, 864=324$ non-isomorphic choices on $X$.
Without loss of generality, assume only $c(ab)=a$ and $c(ad)=a$ hold (which happens for $\frac{1}{8}\,864=108$ non-isomorphic choices on $X$). 
According to cases (1.i), (1.ii), (2.i), and (2.ii), WWARP fails if and only if at least one of the conditions $c(abd) \in bd$, $c(abe)=b$ or $c(ade)=d$ are true.
This happens for\vs\vs\vs 
\begin{center}\small
	$\displaystyle \left(1 - \frac{1}{3} \left(\frac{2}{3}\right)^{\!2}\right) 108 = 92$\vs\vs
\end{center}
 choices.
The same reasoning applies  when only $c(ab)=a$ and $c(ae)=a$ are true, or  only $c(ad)=a$ and $c(ae)=a$ hold. 
Thus, we get a total of $276$ non-WWARP choices.
 
\item[(III)]
All of $c(ab)=a, c(ad)=a, c(ae)=a$ hold.
This happens for a total of $\frac{1}{8}\, 864=108$ non-isomorphic choices on $X$.
According to cases (1.i), (1.ii), (2.i), (2.ii), (3.i), and (3.ii), WWARP fails if and only if at least one of conditions $c(abd) \in bd$, $c(abe) \in be$, or $c(ade) \in de$ holds.
Thus, we get a total of\vs\vs 
\begin{center} \small
	$\displaystyle \left(1 - \left(\frac{1}{3}\right)^{\!3}\right) 108 = 104$\vs\vs
\end{center}
non-WWARP choices on $X$.

\item[(IV)] 
None of $c(ab)=a$, $c(ad)=a$, and $c(ae)=a$ holds. This choice satisfies WWARP.
\end{description}

Since cases (I), (II), (III), and (IV) are mutually exclusive, we conclude that WWARP fails for $180+276+104=560$ choices.
Thus, the number of non-isomorphic WWARP choices on $X$ is $864-560=304$.
\end{proof}

%%%%%%%%%%%%%%%%%%%%%%%%%%%%%%%%%%%
%%%%%%%%%%%%%%%%%%%%%%%%%%%%%%%%%%%

\subsection{Choice with limited attention (CLA)}

{\large \begin{lemma} \label{LEM:CLA}
There are exactly $324$ non-isomorphic CLA choices on $X$.
\end{lemma}}

As announced, instead of giving a formal proof, we present two \textsc{Matlab} programs, which are based on two equivalent formulations of WARP(LA), described in Lemma~\ref{LEM:WARPLA_alternative_formulation}.
The final numbers of CLA choices obtained by running the two different programs are the same, namely 324. 

%To start, we introduce the notion of a \textsl{switch}.

\begin{definition}\label{DEF:minimal_violations_of_alpha}
	For any choice $c \colon \X \to X$, a (minimal) \textsl{switch} is an ordered pair $(A,B)$ of menus such that $A \subseteq B$, $c(A) \neq c(B) \in A$, and $\vert B \setminus A \vert = 1$. 
%Whenever clear from context, we shall simply refer to $\langle A,A \cup \{x\}\rangle$ as a \textsl{violation of} $\alpha$. 
Equivalently, a switch is a pair $(B \setminus x,B)$ of menus such that $c(B \setminus x) \neq c(B) \neq x$.
%\footnote{Recall that $A \cup x$ is an abbreviation for $A \cup \{x\}$.}
%We shall write $A \leftrightharpoons Ax$ to mean that $(A,A \cup \{x\})$ is a minimal switch. 
\end{definition} 

\begin{lemma}\label{LEM:WARPLA_alternative_formulation}
	The following statements are equivalent for a choice $c$:
	\begin{itemize}\vspace{-0.2cm}
		\item[\rm(i)] WARP(LA) holds;\vspace{-0.2cm}
		\item[\rm(ii)] for any $A\in\X$, there is $x\in A$ such that, for any $B$ containing $x$, if $c(B)\in A$, then $(B\setminus x,B)$ is not a switch;\vspace{-0.2cm} 
		\item[\rm(iii)] there is a linear order $>$ on $X$ such that, for any $x,y\in X$, $x>y$ implies that there is no switch $(B\setminus y,B)$ such that $c(B)=x$.
		\end{itemize}
\end{lemma}

\noindent \textit{Proof of Lemma~\ref{LEM:WARPLA_alternative_formulation}.} 
%WARP(LA) can be restated as follows:
The equivalence between (i) and (ii) follows from the definition of WARP(LA) and Definition \ref{DEF:minimal_violations_of_alpha}.
To show that (iii) implies (ii), for any $A\in \X$, take $x:=\min(A,>)$.
To show that (ii) implies (iii), assume property (ii) holds.
Thus, for $A:=X$, there is $x\in X$ such that, for any $B$ containing $x$, $(B\setminus x,B)$ is not a switch.
Next, let $A:=X\setminus x$.
By (ii), there is $x^{\prime}\in X\setminus x$ such that, for any $B$ containing $x^{\prime}$, if $c(B)\in X\setminus x$ (equivalently, $c(B)\neq x$), then $(B\setminus x^{\prime},B)$ is not a switch.
Set $x>x^{\prime}$, and take $A:=X\setminus xx^{\prime}$.
By (ii),  there is $x^{\prime\prime}\in X\setminus xx^{\prime}$ such that, for any $B$ containing $x^{\prime\prime}$, if $c(B)\in X\setminus xx^{\prime}$ (equivalently, $c(B)\neq x,x^{\prime}$), then $(B\setminus x^{\prime\prime},B)$ is not a switch.
Set $x> x^{\prime\prime}$ and $x^{\prime}>x^{\prime\prime}$.
Thus, we get the transitive chain $x > x' > x''$.
Since $X$ is finite, we can continue this process until obtaining what we are after. \qed

\bigskip

To compute the number of non-isomorphic choices on $X=abde$ on \textsc{MATLAB}, we first list all $864$ non-isomorphic choice functions satisfying $c(abde)=e$, $c(abd)=d$, and $c(ab)=b$.\footnote{{} This is equivalent to requiring $c(abde)=a$, $c(bde)=b$, and $c(de)=d$, as in Lemma \ref{LEM:WWARP}.}
In the code, we set $a:=\,$\texttt{1}, $b:=\,$\texttt{2}, $d:=\,$\texttt{3}, and $e:=\,$\texttt{4}. 
Moreover, each subset of $abde:=\,$\texttt{1234} is labeled by a number, which goes from \texttt{1} to \texttt{11}. 
(Since we do not consider singletons and the empty set, there are only 11 feasible menus.)

\begin{lstlisting}
pkg load communications 
%%%%%%%%%%%%%%%%%%%%%%% PRINT A CHOICE IN A NICE FORM %%%%%%%%%%%%%%%%%%%%%%% 

function y = listofallchoicesiso()
% a=1, b=2, c=3, and d=4.
% Generating all choices such that c(1234) = 4, c(123) = 3, c(12) = 2.
% Encoding of sets: 12 is 1, 13 is 2, 23 is 3, 14 is 4, 24 is 5, 34 is 6, 123 is 7, 124 is 8, 134 is 9, 234 is 10, and 1234 is 11.

y = [];
for a = [1,3]  
for b = [2,3]  
for c = [1,4]  
for d = [2,4]  
for e = [3,4] 
for f = [1,2,4]
for g = [1,3,4]
for h = [2,3,4]    
choice(1) = 2;
choice(2) = a;
choice(3) = b;
choice(4) = c;
choice(5) = d;
choice(6) = e;
choice(7) = 3;
choice(8) = f;
choice(9) = g; 
choice(10) = h;
choice(11) = 4;
y = [y;choice];
end
end
end
end
end
end
end
end
end
%%%%%%%%%%%%%%%%%%%%%%%%%%%%%%%%%%%%%%%%%%%%%%%%%%%%%%%%%%%%%%%%%%%%%%%%%%%%%

\end{lstlisting}

We build a function, called \texttt{index2array(x)}, which displays, for any menu A (denoted by \texttt{x} in the code), the array of its elements.

\begin{lstlisting}
%%%%%%%%%%%%%%%%%%%%%%%%%%%% DISPLAY ALL MENUS %%%%%%%%%%%%%%%%%%%%%%%%%%%%%%
	function y = index2array(x)
% Takes as input the integer representing a set, and returns the array
% of its elements.

if (x == 1)
  y = [1,2];
elseif (x == 2)
  y = [1,3];
elseif (x == 4)
  y = [1,4];
elseif (x == 3)
  y = [2,3];
elseif (x == 5)
  y = [2,4];
elseif (x == 6)
  y = [3,4];
elseif (x == 7)
  y = [1,2,3];
elseif (x == 8)
  y = [1,2,4];
elseif (x == 9)
  y = [1,3,4];
elseif (x == 10)
  y = [2,3,4];
elseif (x == 11)
  y = [1,2,3,4];
else
  disp('not found');
endif
end
%%%%%%%%%%%%%%%%%%%%%%%%%%%%%%%%%%%%%%%%%%%%%%%%%%%%%%%%%%%%%%%%%%%%%%%%%%%%%
\end{lstlisting}

Next, the function \texttt{listswitches(x)} takes as input a choice $c$ (denoted by \texttt{x} in the code) on $X=abde$, and lists as output all the switches of $c$. 
The list \textsl{switches} includes all possible switches of a choice function.
Note that each switch $(B\setminus x,B)$ is encoded as \texttt{[p,q,r]}, meaning that \texttt{p}$=c(B\setminus x)$, \texttt{q}$=c(B)$, and \texttt{r}$=x$.
The function \textsl{switches} returns the 3-column matrix of all switches.
Each row displays a switch in the form discussed above. 
\begin{lstlisting}
%%%%%%%%%%%%%%%%%%%%%%%%%%%%%% LIST ALL SWITCHES %%%%%%%%%%%%%%%%%%%%%%%%%%%%
function y = listswitches(x)
% This creates set of all switches:  p is chosen in a menu, q is added to it, and now r is chosen from the enlarged menu (p,q,r are all different).
% Each switch is between a menu of size 2 and a menu of size 3, or else between a menu of size 3 and the ground set 1234.  
% In this function, the input is a choice function (denoted by x) on the ground set 1234. 
% Each choice is an array mapping sets to points. 
% We ignore singletons.
  
switches = [];  % This is the array of all switches. 
if (x(1) == 1 && x(7) == 2)  
% This means that the first set (1,2) chooses 1, and the seventh set (1,2,3) chooses 2. 
% This is a switch.

      switches = [switches;[1,2,3]];
endif
if (x(1) == 2 && x(7) == 1)
      switches = [switches;[2,1,3]];
endif
if (x(1) == 1 && x(8) == 2)
      switches = [switches;[1,2,4]];
endif
if (x(1) == 2 && x(8) == 1)
      switches = [switches;[2,1,4]];
endif
if (x(2) == 1 && x(7) == 3)
      switches = [switches;[1,3,2]];
endif
if (x(2) == 3 && x(7) == 1)
      switches = [switches;[3,1,2]];
endif
if (x(2) == 1 && x(9) == 3)
      switches = [switches;[1,3,4]];
endif
if (x(2) == 3 && x(9) == 1)
      switches = [switches;[3,1,4]];
endif
if (x(3) == 2 && x(7) == 3)
      switches = [switches;[2,3,1]];
endif
if (x(3) == 3 && x(7) == 2)
      switches = [switches;[3,2,1]];
endif
if (x(3) == 2 && x(10) == 3)
      switches = [switches;[2,3,4]];
endif
if (x(3) == 3 && x(10) == 2)
      switches = [switches;[3,2,4]];
endif
if (x(4) == 1 && x(8) == 4)
      switches = [switches;[1,4,2]];
endif
if (x(4) == 4 && x(8) == 1)
      switches = [switches;[4,1,2]];
endif
if (x(4) == 1 && x(9) == 4)
      switches = [switches;[1,4,3]];
endif
if (x(4) == 4 && x(9) == 1)
      switches = [switches;[4,1,3]];
endif
if (x(5) == 2 && x(8) == 4)
      switches = [switches;[2,4,1]];
endif
if (x(5) == 4 && x(8) == 2)
      switches = [switches;[4,2,1]];
endif
if (x(5) == 2 && x(10) == 4)
      switches = [switches;[2,4,3]];
endif
if (x(5) == 4 && x(10) == 2)
      switches = [switches;[4,2,3]];
endif
if (x(6) == 3 && x(9) == 4)
      switches = [switches;[3,4,1]];
endif
if (x(6) == 4 && x(9) == 3)
      switches = [switches;[4,3,1]];
endif
if (x(6) == 3 && x(10) == 4)
      switches = [switches;[3,4,2]];
endif
if (x(6) == 4 && x(10) == 3)
      switches = [switches;[4,3,2]];
endif
% These are all switches between menus with 2 items and menus with 3 items. 
% The rest are all switches between menus with 3 items and the ground set X. 
if (x(7) == 1 && x(11) == 2)
      switches = [switches;[1,2,4]];
endif
if (x(7) == 1 && x(11) == 3)
      switches = [switches;[1,3,4]];
endif
if (x(7) == 2 && x(11) == 1)
      switches = [switches;[2,1,4]];
endif
if (x(7) == 2 && x(11) == 3)
      switches = [switches;[2,3,4]];
endif
if (x(7) == 3 && x(11) == 1)
      switches = [switches;[3,1,4]];
endif
if (x(7) == 3 && x(11) == 2)
      switches = [switches;[3,2,4]];
endif
if (x(8) == 1 && x(11) == 2)
      switches = [switches;[1,2,3]];
endif
if (x(8) == 1 && x(11) == 4)
      switches = [switches;[1,4,3]];
endif
if (x(8) == 2 && x(11) == 1)
      switches = [switches;[2,1,3]];
endif
if (x(8) == 2 && x(11) == 4)
      switches = [switches;[2,4,3]];
endif
if (x(8) == 4 && x(11) == 1)
      switches = [switches;[4,1,3]];
endif
if (x(8) == 4 && x(11) == 2)
      switches = [switches;[4,2,3]];
endif
if (x(9) == 1 && x(11) == 3)
      switches = [switches;[1,3,2]];
endif
if (x(9) == 1 && x(11) == 4)
      switches = [switches;[1,4,2]];
endif
if (x(9) == 3 && x(11) == 1)
      switches = [switches;[3,1,2]];
endif
if (x(9) == 3 && x(11) == 4)
      switches = [switches;[3,4,2]];
endif
if (x(9) == 4 && x(11) == 1)
      switches = [switches;[4,1,2]];
endif
if (x(9) == 4 && x(11) == 3)
      switches = [switches;[4,3,2]];
endif
if (x(10) == 2 && x(11) == 3)
      switches = [switches;[2,3,1]];
endif
if (x(10) == 2 && x(11) == 4)
      switches = [switches;[2,4,1]];
endif
if (x(10) == 3 && x(11) == 2)
      switches = [switches;[3,2,1]];
endif
if (x(10) == 3 && x(11) == 4)
      switches = [switches;[3,4,1]];
endif
if (x(10) == 4 && x(11) == 2)
      switches = [switches;[4,2,1]];
endif
if (x(10) == 4 && x(11) == 3)
      switches = [switches;[4,3,1]];
endif
y = switches; 
% The function returns the array of all switches. 
% It is a matrix. 
% Each row is a switch (3 items). 
end      
%%%%%%%%%%%%%%%%%%%%%%%%%%%%%%%%%%%%%%%%%%%%%%%%%%%%%%%%%%%%%%%%%%%%%%%%%%%%%   
\end{lstlisting}

The following function, named \texttt{secontainselement(z)}, checks whether an item $x$ belongs to some set $A\in \X$.
In the code the object \texttt{z} denotes a pair consisting of an item, denoted by \texttt{z(1)}, and a menu, denoted by \texttt{z(2)}.
The function returns \texttt{1} if \texttt{z(1)} belongs to \texttt{z(2)}, and \texttt{0} otherwise.
This function will be used to test the alternatives formulations of WARP(LA) described in Lemma \ref{LEM:WARPLA_alternative_formulation}. 

\begin{lstlisting}
%%%%%%%%%%%%%%%%%%%%%%%%%%%%%%%%%%%%%%%%%%%%%%%%%%%%%%%%%%%%%%%%%%%%%%%%%%%%% 	
	function y = setcontainselement(z) 
	% ordered pair, first item x then set p  
x = z(1); % The input z is a pair, consisting of an item x and a set p (which is encoded as an integer)
p = z(2);
if ((p == 1 && x == 1) || (p == 1 && x == 2) || (p == 2 && x == 1) || (p == 2 && x == 3)) 
  y = 1;  % p=1 && x=2 means the 1st set (1,2) and the 2nd element 2, since 2 in (1,2) we return y = 1, which means true.
elseif ((p == 3 && x == 2) || (p == 3 && x == 3) || (p == 4 && x == 1) || (p == 4 && x == 4))
  y = 1;
elseif ((p == 5 && x == 2) || (p == 5 && x == 4) || (p == 6 && x == 3) || (p == 6 && x == 4))
y = 1;
elseif ((p == 7 && x == 1)||(p == 7 && x == 2)||(p == 7 && x == 3)||(p == 8 && x == 1)||(p == 8 && x == 2))
y = 1;
elseif ((p == 8 && x == 4)||(p == 9 && x == 1)||(p == 9 && x == 3)||(p == 9 && x == 4)||(p == 10 && x == 2)) 
y = 1;
elseif ((p == 10 && x == 3)||(p == 10 && x == 4)||(p == 11 && x == 1)||(p == 11 && x == 2))
y = 1;
elseif ((p == 11 && x == 3)||(p == 11 && x == 4))
y = 1;
else
y = 0; % If the point is not in the set, we return 0 for false.
endif 
end
%%%%%%%%%%%%%%%%%%%%%%%%%%%%%%%%%%%%%%%%%%%%%%%%%%%%%%%%%%%%%%%%%%%%%%%%%%%%%
\end{lstlisting}
The next code counts the number of non-isomorphic choice functions on $X$ satisfying the property described in Lemma \ref{LEM:WARPLA_alternative_formulation} (ii).
The function \texttt{prelimtestWARPLA(A,S,x)}, for any choice function $c$, takes as input a set $A\in X$ (denoted by \texttt{A}), the family of all switches of $c$ (represented by the matrix \texttt{S}), and an item $x\in X$ (denoted by \texttt{x}), and checks whether there is a switch $(B,B\setminus x)$ such that $c(B)\in A$. 
This function gives \texttt{0} if such a switch exists, otherwise returns \texttt{1}.
Thus, WARP(LA) can be restated as  for all nonempty \texttt{A} there exists \texttt{x} $\in$ \texttt{A} such that the  function \texttt{prelimtestWARPLA(A,S,x)} returns \texttt{1} on input \texttt{(A,S,x)} where \texttt{S} is the list of all existing switches.

The function \texttt{testifAisWARPLA(A,S)}, for a given choice $c$, takes as input a menu $A$ (denoted by \texttt{A}) and the family of all switches of $c$ (described in the matrix \texttt{S}), and test whether there is $x\in A$ such that $(B,B\setminus x)$ is a switch and $c(B)\in A$.
This function uses \texttt{setcontainselement(m)}, \texttt{index2array(A)}, and \texttt{prelimtestWARPLA(A,S,x)}, which were previously built, and gives \texttt{1} if it finds some $x$ satisfying the required constraints, or \texttt{0} otherwise.

The function \texttt{testifchoiceisWARPLA(x)} takes as input a choice function $c$ (denoted by \texttt{x}) and, testing all the menus of $c$ using \texttt{testifAisWARPLA(A,S)}, returns \texttt{1} if $c$ satisfies WARP(LA), and \texttt{0} otherwise.

We collect all the choices satisfying WARPLA in the list \texttt{WARPLA}, while we put the other choices in the list \texttt{notWARPLA}, and we display, using the commands 
	\texttt{size(WARPLA)} and \texttt{size(notWARPLA)}, the size of these lists, obtaining what we are looking for.  
\begin{lstlisting}
%%%%%%%%%%%%%%%%%%%%%%%%%%%%%%%%% METHOD 1 %%%%%%%%%%%%%%%%%%%%%%%%%%%%%%%%%%

% For method 1, we use the characterization stated in Lemma 11(ii).

function y = prelimtestWARPLA(A,S,x) 
% [A,S,x] for a point x, set A (1 to 11), and set of switches S
% This first function takes as input the set of switches, an item x and 
% a set A, and checks if there is any switch ( , in A, x).
s = size(S)(1);
for j = 1:s  % Looping through all switches
       m = [S(j,2),A]; 
    if (S(j,3) == x && setcontainselement(m) == 1) % Add x and c(B) in A
          y = 0;  % If so, then we have a switch (, in A, x) 
          return;
    endif   
end
y = 1; % Otherwise return 1 because we didn't find such a switch.
end

% Using the function above, we can restate WARPLA as 
% for any menu A, there is x in A such that this function returns 1 
% on input (A,S,x), where S is the list of switches.

% The next function searches for each A whether there is an item x or not.
function q = testifAisWARPLA(A,S) %[A,S] for a given set A (1 to 11) and set of switches S  
B = index2array(A);
n = size(B)(2);  % The set A is now an array of length n, the size of  A
for i = 1:n    
  z = prelimtestWARPLA(A,S,B(i)); % Thus B(i) is looping over the items of A
  if (z == 1)  % As soon as we find a x that works we return a 1
    q = 1;
  return;
  endif
endfor
q = 0; % If we cannot find an item x that works, then we return 0.
end  

% Using the function above, we can restate WARP(LA) 
% as for all nonempty A, the function above returns 1 on input (A,S), where S
% is the set of switches.

function y = testifchoiceisWARPLA(x) %%%%%%%%%%%%%%%%%%%%%%%%%%%%%%%%%%%%%%%
S = listswitches(x);
for i = 1:11;  % Testing all A
  j = testifAisWARPLA(i,S);
  if (j == 0)
    y = 0;
    return
  endif
y = 1;
end
end

function testWARPLA %%%%%%%%%%%%%%%%%%%%%%%%%%%%%%%%%%%%%%%%%%%%%%%%%%%%%%%%
y = listofallchoicesiso();% Here we loop through all choice functions
WARPLA = []; % We list WARP(LA) choice functions 
notWARPLA = []; % We list those that are not WARP(LA)
for i = 1:864
  x = y(i,:); % x is the i-th choice function
  if testifchoiceisWARPLA(x) == 1
    WARPLA = [WARPLA;x];
  else
    notWARPLA = [notWARPLA;x];
  endif
end  
disp('number of WARPLA is: ')
size(WARPLA)(1) % Printing the number of choices that are WARP(LA)
disp('number of NOT WARPLA is: ')
size(notWARPLA)(1) % Printing the number of choices that are not WARP(LA)
end

%%%%%%%%%%%%%%%%%%%%%%%%%%%%%%%%%%%%%%%%%%%%%%%%%%%%%%%%%%%%%%%%%%%%%%%%%%%%%  	
\end{lstlisting}

In the following code, we compute the number of choices satisfying the property stated in Lemma \ref{LEM:WARPLA_alternative_formulation}(iii).
We need to check whether, given a choice $c$ and the associated switches, a linear order $>$ on $X$ satisfies
\begin{equation}\label{EQ:switches}
x>y \quad \Longrightarrow \quad \Big(c(B)=x \quad \Longrightarrow \quad \big(c(B)=c(B\setminus y)\; \vee\; c(B)=y \big)\Big)
\end{equation}
for any $x,y\in X$ and $B\in\X$ containing $x,y$.
To that end,  we first build the function \texttt{testifsetofswitchesisorderablebyperm(S,q)}, which takes as inputs the  family of all switches (represented on \textsc{Matlab} by the matrix \texttt{S}) of a given choice function $c$, and a given linear order $>$ on $X$ (represented by a permutation $q$ of the set 1234), and returns \texttt{0} if $>$ satisfies Condition \ref{EQ:switches}, or \texttt{1} otherwise.

The function \texttt{perms([1,2,3,4])} generates all the linear orders on $X$ (i.e. all the possible permutations of the set 1234).
The function  \texttt{testswitchesWARPLA(S)} takes as input the family of all switches of a choice function $c$, and returns \texttt{1} if there is a linear order $>$ satisfying Condition \ref{EQ:switches}, and \texttt{0} otherwise.
Finally, we define the function \texttt{testWARPLA2}.
This command first checks, for any choice $c$ (which is denoted by \texttt{x} in \textsc{Matlab}), whether it satisfies the property stated in Lemma~\ref{LEM:WARPLA_alternative_formulation}(iii).
Then  the function collects the choices satisfying the alternative formulation of WARP(LA) in the list \texttt{in}, and the other choices in the list \texttt{out}, and displays the size of these lists, obtaining the number of non-isomorphic choices satisfying WARP(LA) (and the number of those which do not satisfy it).    

\begin{lstlisting}
%%%%%%%%%%%%%%%%%%%%%%%%%%%%%%% METHOD 2 %%%%%%%%%%%%%%%%%%%%%%%%%%%%%%%%%%%%

% In this method, we use the following characterization of WARPLA:
% there is an total ordering < of the underlying set X such that there is
% no switch ( , x, y) with x < y.

% So the first thing we do is check whether a given ordering, given as 
% a permutation of (1,2,3,4), works or not

function y = testifsetofswitchesisorderablebyperm(S,q)  %%%%%%%%%%%%%%%%%%%
M = size(S)(1); % Number of switches
for m=1:M % Looping through switches
    if (q(S(m,2)) < q(S(m,3))) % Checks whether y<x and c(B)=x  c(B\y).
       y = 0; % If y=0, then the ordering does not work
       return
    endif
end
y = 1; % Otherwise returns 1, because we found the desired ordering
end

function y = testswitchesWARPLA(S) %%%%%%%%%%%%%%%%%%%%%%%%%%%%%%%%%%%%%%%%%
P = perms([1,2,3,4]); % Here we generate all orderings (permutations)
for (n = 1:24) % There are 24 permutations of 4 elements, we loop through them
    if testifsetofswitchesisorderablebyperm(S,P(n,:)) % Checks the n-th order 
       y = 1; % If it worked, then it is WARPLA
       return
   endif  
end  
y = 0; % If no permutation worked, then it is not WARPLA
end
  

  
function testWARPLA2  %%%%%%%%%%%%%%%%%%%%%%%%%%%%%%%%%%%%%%%%%%%%%%%%%%%%%%%
y = listofallchoicesiso(); 
in = []; % List of WARP(LA) choices
out = []; % List of not WARP(LA) choices
for i = 1:864
  x = y(i,:); % Check choice x
  if testswitchesWARPLA(listswitches(x)) == 1
    in  = [in;x]; % x is WARPLA
  else
    out = [out;x];
  endif
end  
disp('number of WARPLA here is: ')
size(in)(1)
disp('number of NOT WARPLA is: ')
size(out)(1)
end  
%%%%%%%%%%%%%%%%%%%%%%%%%%%%%%%%%%%%%%%%%%%%%%%%%%%%%%%%%%%%%%%%%%%%%%%%%%%%%
\end{lstlisting}

We run \texttt{testWARPLA} and \texttt{WARPLA2}.
There are 324 non-isomorphic choices on $X$ satisfying property (ii) in Lemma~\ref{LEM:WARPLA_alternative_formulation}.
The same number of choices satisfies property (iii) of the mentioned result.
We conclude that the number of non-isomorphic CLA choice on $X$ is $324$. 
\medskip

\begin{lstlisting}
testWARPLA
testWARPLA2	
\end{lstlisting} 

\begin{table*}
	
\end{table*}

\begin{verbatim}

-----------------------------------------------------------------------------	
number of WARPLA is: 
ans =  324
number of NOT WARPLA is: 
ans =  540
number of WARPLA here is: 
ans =  324
number of NOT WARPLA is: 
ans =  540
---------------------------------------------------------------------------
\end{verbatim} 
\section*{Declaration of competing interests}
The authors declare that they have no known competing financial interests or personal relationships that could have appeared to influence the work reported in this paper.

%%%%%%%%%%%%%%%%%%%%%%%%%%%%%%
%%%%%%%%%%%%%%%%%%%%%%%%%%%%%%
%%%%%%%%%%%%%%%%%%%%%%%%%%%%%%
%%%%%%%%%%%%%%%%%%%%%%%%%%%%%%


\begin{thebibliography}{31}

\providecommand{\natexlab}[1]{#1}
\providecommand{\url}[1]{\texttt{#1}}
\expandafter\ifx\csname urlstyle\endcsname\relax 
  \providecommand{\doi}[1]{doi: #1}\else
  \providecommand{\doi}{doi: \begingroup \urlstyle{rm}\Url}\fi

%\bibitem[Afriat(1974)]{Afriat1974}
%{\textsc{Afriat, S.} 1974.
%On a system of inequalities in demand analysis: an extension of the classical method.
%\textit{International Economic Review} 14:\,460--472.}

\bibitem[Apesteguia and Ballester(2013)]{ApesteguiaBallester2013}
{\textsc{Apesteguia, J., and Ballester, M.\,A.}, 2013.
Choice by sequential procedures.
\textit{Games and Economic Behavior} 77:\,90--99.} 

%\bibitem[Apesteguia and Ballester(2017)]{ApesteguiaBallester2017}
%{\textsc{Apesteguia, J., and Ballester, M.\,A.}, 2017.
%A measure of rationality and welfare.
%\textit{Journal of Political Economy} 123:\,1278--1310.}
% 
%\bibitem[Cantone, Giarlotta, and Watson(2019)]{CantoneGiarlottaWatson2019}{\textsc{Cantone, D., Giarlotta, A., and Watson, S.}, 2019.
%Congruence relations on a choice space.  
%\textit{Social Choice and Welfare} 52: 247--294.}
%
%\bibitem[Cantone, Giarlotta, and Watson(2021)]{CantoneGiarlottaWatson2021}
%{\textsc{Cantone, D., Giarlotta, A., and Watson, S.}, 2021.
%Choice resolutions.   
%\textit{Social Choice and Welfare} 56:\,713--753.} 

\bibitem[Cherepanov, Feddersen, and Sandroni(2013)]{CherepanovFeddersenSandroni2013}
{\textsc{Cherepanov, V., Feddersen, T., and Sandroni, A.}, 2013.
Rationalization. 
\textit{Theoretical Economics} 8:\,775--800.}

%\bibitem[Chernoff(1954)]{Chernoff1954}
%{\textsc{Chernoff, H.}, 1954. 
%Rational selection of decision functions. 
%\textit{Econometrica} 22:\,422--443.}

%\bibitem[de Clippel and Rozen(2021)]{DeClippleRozen2021}{\textsc{de Clippel, G., and Rozen, K.}, 2021.
%Bounded rationality and limited datasets.
%\textit{Theoretical Economics}
%16:\,359--380.}
%
%\bibitem[Garc\'ia-Sanz and Alcantud(2015)]{GarciaAlcantud2015}
%{\textsc{Garc\'ia-Sanz, M., and Alcantud J.\,C.\,R.}, 2015.
%Sequential rationalization of multivalued choice.
%\textit{Mathematical Social Sciences} 74:\,29--33.}
 
\bibitem[Giarlotta, Petralia, and Watson(2022)]{GiarlottaPetraliaWatson2021}{\textsc{Giarlotta, A., Petralia, A., and Watson, S.}, 2022.
Bounded rationality is rare.
\textit{Journal of Economic Theory}, forthcoming.} 

\bibitem[Lleras, Masatlioglu, Nakajima, and Ozbay(2017)]{LlerasMasatliogluNakajimaOzbay2017}
{\textsc{Lleras, J.,\, S, Masatlioglu, Y., Nakajima, D., and Ozbay, E.\,Y.}, 2017.
When more is less: limited consideration.
\textit{Journal of Economic Theory} 170:\,70--85.} 

\bibitem [Manzini and Mariotti(2007)]{ManziniMariotti2007} 
{\textsc{Manzini, P., and Mariotti, M.}, 2007.
Sequentially rationalizable choice. 
\textit{American Economic Review} 97:\,1824--1839.} 

\bibitem [Manzini and Mariotti(2012a)]{ManziniMariotti2012a}
{\textsc{Manzini, P., and Mariotti, M.}, 2012a.
Choice by lexicographic semiorders.  
\textit{Theoretical Economics} 7:\,1--23.} 

\bibitem [Manzini and Mariotti(2012b)]{ManziniMariotti2012b}
{\textsc{Manzini, P., and Mariotti, M.}, 2012b.
Categorize then choose: Boundedly rational choice and welfare.  
\textit{Journal of the European Economic Association} 10:\,1141--1165.}

\bibitem[Masatlioglu, Nakajima, and Ozbay(2012)]{MasatliogluNakajimaOzbay2012}
{\textsc{Masatlioglu, Y., Nakajima, D., and Ozbay, E.\,Y.}, 2012.
Revealed attention. 
\textit{American Economic Review} 102:\,2183--2205.}

%\bibitem[Masatlioglu and Ok(2005)]{MasatliogluOk2005}
%{\textsc{Masatlioglu, Y., and Ok, E.\,A.}, 2005.
%Rational choice with status quo bias. 
%\textit{Journal of Economic Theory} 121:\,1--29.}
%
\bibitem[Rubinstein and Salant(2008)]{RubinsteinSalant2008}
{\textsc{Rubinstein, A., and Salant, Y.}, 2008.
Some thoughts on the principle of revealed preference.
In: Schotter, A., and Caplin, A.,
\textit{The Foundations of Positive and Normative Economics: A Handbook} pp.\,116--124. Oxford University Press.} 

\bibitem[Samuelson(1938)]{Samuelson1938}
{\textsc{Samuelson, A.\,P.}, 1938.
A note on the pure theory of consumer's behaviour.
\textit{Economica} 17:\,61--71.} 

%\bibitem[Selten(1991)]{Selten1991}
%{\textsc{Selten, R.}, 1991.
%Properties of a measure of predictive success.
%\textit{Mathematical Social Sciences} 21:\,153--167.}

\bibitem[Simon(1955)]{Simon1955}
{\textsc{Simon, H.\,A.}, 1955.
A behavioral model of rational choice.
\textit{Quarterly Journal of Economics} 69:\,99--118.}

\bibitem[Xu and Zhou(2007)]{XuZhou2007}
{\textsc{Xu, Y., and Zhou, L.}, 2007.
Rationalizability of choice functions by game trees.
\textit{Journal of Economic Theory} 134:\,548--556.} 

\bibitem[Yildiz(2016)]{Yildiz2016}
{\textsc{Yildiz, K.}, 2016.
List-rationalizable choice.
\textit{Theoretical Economics} 11:\,587--599.}

\end{thebibliography}
\end{document}